\documentstyle[11pt,dina4p,epic,twoside,epsfig]{article}
\begin{document}

\thispagestyle{empty}

\hspace*{11cm}{ \Large HEP-EX/9507013 }

\vspace*{2cm}
\begin{center}
  \huge An Investigation into the Radiation Damage\\
        of the Silicon Detectors of the H1-PLUG Calorimeter\\
        within the HERA environment
\end{center}
\vspace*{2cm}

\begin{center}
  \large W. Hildesheim\footnote{Deutsches Elektronen Synchrotron,
    Hamburg}, M. Seidel\footnote{I. Institut f\"ur Experimentalphysik,
    Universit\"at Hamburg}
\end{center}

\vspace*{4cm}

\begin{abstract}

  The silicon detectors used in the H1-PLUG calorimeter have shown
  increasing aging effects during the '94 run period of the electron
  proton storage ring HERA.  These effects were particularly manifest
  as degradation of the signal to noise level and the calibration
  stability.  The reasons for this behaviour have been found to be
  correlated with radiation damage to the silicon oxide passivation
  edges of the detectors in strong and fluctuating increases of the
  leakage currents and in severe changes of the flat band voltages.
  Depletion voltages however are found to be stable and therefore bulk
  damage of the silicon can be excluded.  A comparison with
  measurements made by thermoluminescence dosimeters as well as
  related laboratory experiments suggest that the aging is due to very
  low energetic electrons and photons.
\end{abstract}

\clearpage
\newpage

\thispagestyle{empty}
\setcounter{page}{1}
\pagenumbering{arabic}

\tableofcontents

\clearpage

\section{Introduction}

Silicon detectors are being increasingly applied to high energy
physics experiments. Silicon technology has been choosen by
experiments for purposes as different as the construction of vertex
detectors, trackers and calorimeters.  As an example the ATLAS
collaboration \cite{atlas} at the future LHC will instrument the
complete inner part of their detector with silicon. One reason for
this is the high radiation hardness of silicon which is a requirement
for all present and future accelerators.  Two special collaborations
at CERN, RD2 and RD20 currently investigating all problems related to
radiation hardness \cite{rd02}, \cite{rd20}, are focussing on hadronic
radiation damage.

The H1-PLUG-calorimeter is the first hadronic silicon instrumented
sampling calorimeter and has been operational since '92 within the
environment of the HERA electron proton collider. Severe radiation
damage of the silicon detectors used has been observed in particular
during the '94 run period after a strong increase of the delivered
luminosity. The observed radiation damages will be presented and
discussed in this paper.

Following the description of the PLUG calorimeter
(sect.~\ref{section2}), the basic characteristics of the silicon
detectors used are presented (sect.~\ref{section3}). The PLUG readout
circuit is explained (sect.~\ref{PLUGRC}) as an introduction into the
experimental difficulties of the monitoring of the diode
characteristics during normal PLUG operation (sect.~\ref{monit}). The
results of the monitoring during the '94 run period of HERA are given
(sect.~\ref{heram}), and are compared with the final analysis of
related measurements as performed in the laboratory
(sect.~\ref{labom}). The results are finally discussed in
section~\ref{discu}.

\section{PLUG Calorimeter Description}
\label{section2}

The PLUG subdetector of H1 is a silicon instrumented sampling
calorimeter, designed to fill the gap between the forward part of the
LAr calorimeter and the beam pipe and thus to ensure the hermeticity
of the energy measurement \cite{h1det}. In recent applications the
PLUG has proved to play an important role for the tagging of rapidity
gap events \cite{rapgap}.

\begin{figure}[htbp]
  \begin{center}
    \vspace{0cm}
    \leavevmode
    \epsfig{figure=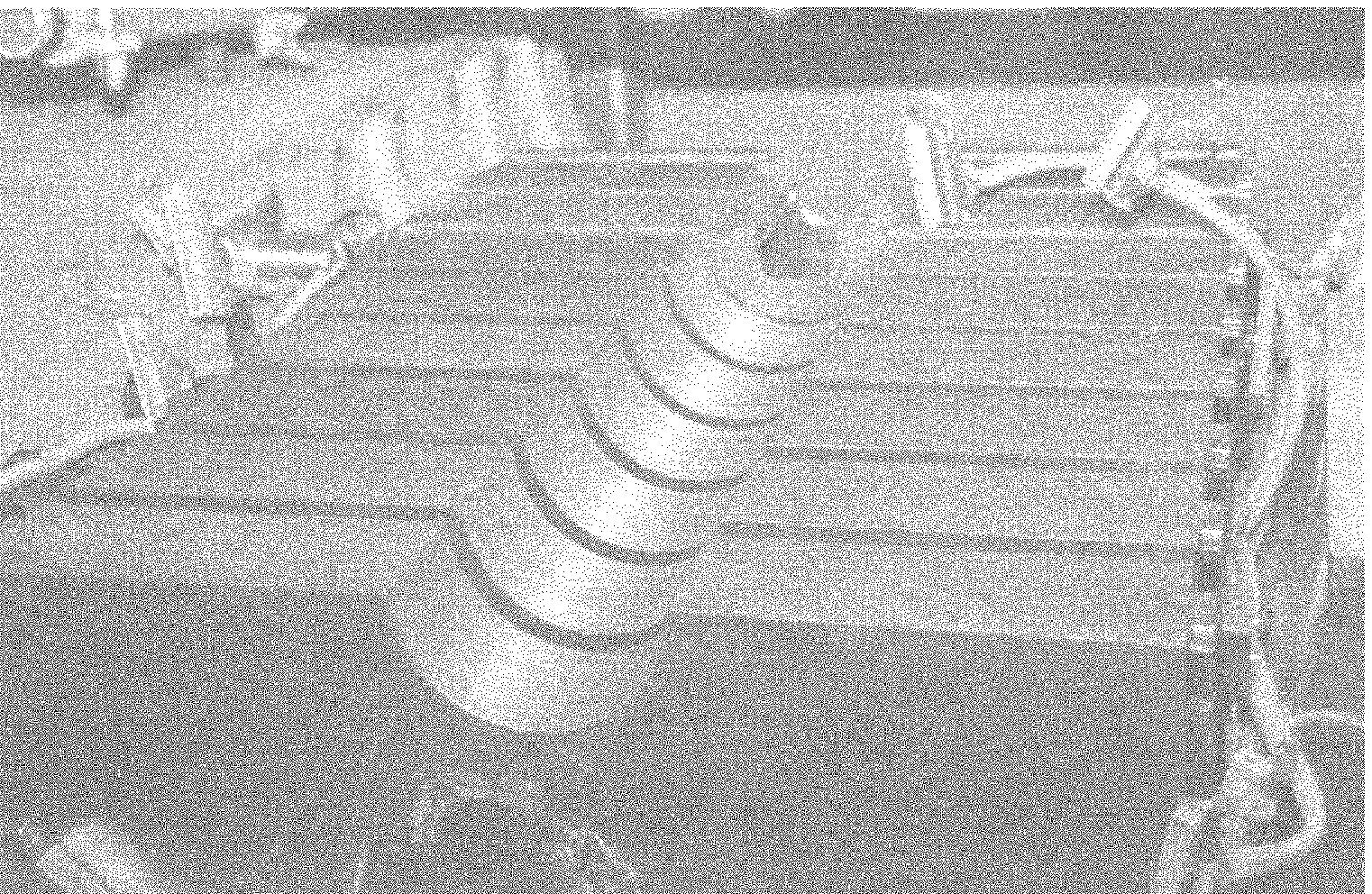,height=10cm,angle=90}
    \hspace{2cm}
    \epsfig{figure=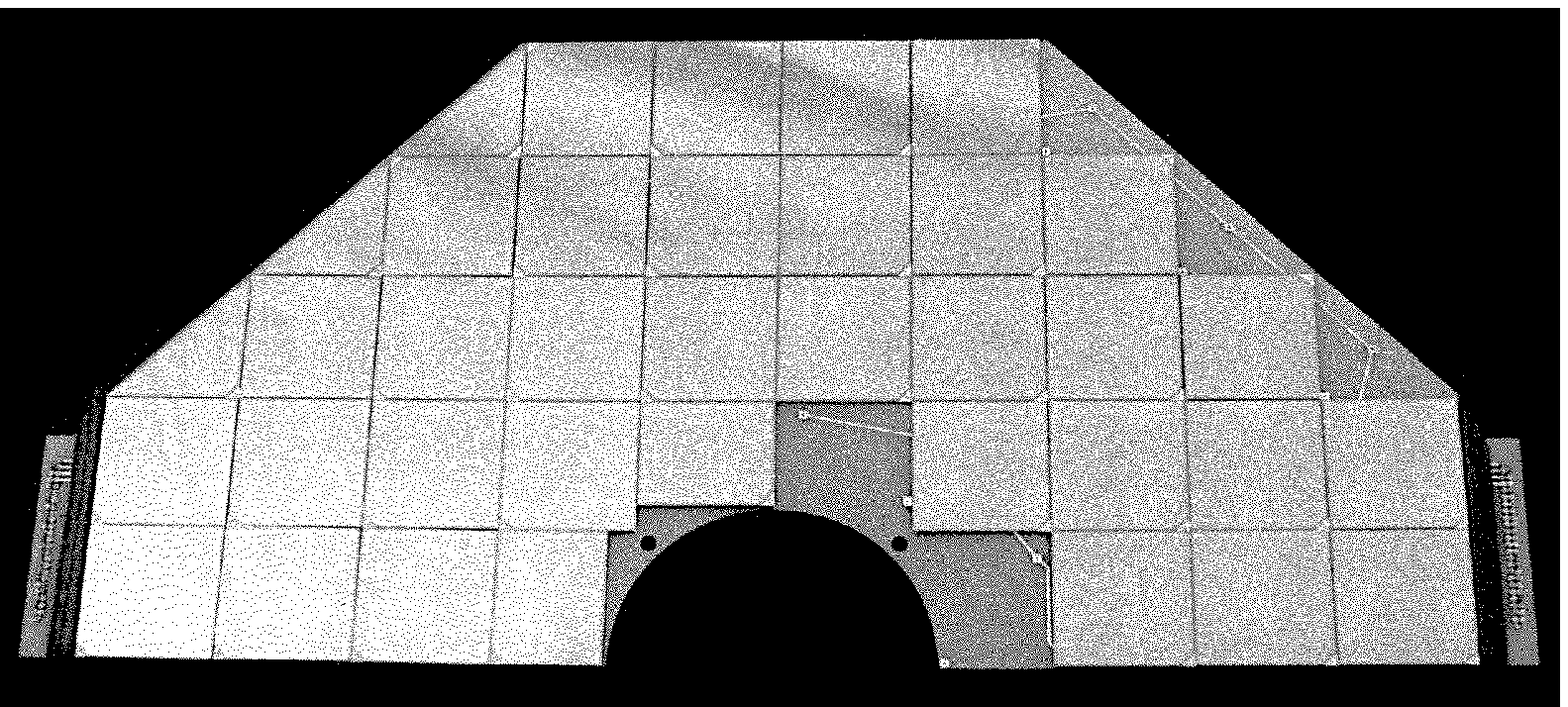,height=10cm,angle=-90}
    \end{center}
  \caption[Half of the PLUG detector]{Half of the PLUG detector is visible
    including eight absorber planes without the instrumented modules.
    One module of the PLUG with 37 installed detectors is shown,  5
    detectors are still missing in order to show a part of the read out
    board. One plane consists of two such modules.}
  \label{fighalf}
\end{figure}

The PLUG is mounted in the return yoke of the magnet, one half is
shown in the left part of figure~\ref{fighalf} with the slots in
between the copper absorbers visible for the 8 detector modules.  The
inner part of such a module is displayed in the right part of figure
\ref{fighalf}. It is instrumented with a total number of 42 detectors
most of which are square shaped (5 x 5 $cm^{2}$) complemented by
rectangular wafers close to the beam pipe and triangular ones at the
outer rim. All detectors have a total thickness of arround 400~$\mu m$
\cite{guide}. The silicon detectors are fabricated in our own
laboratory using a unique combination of a Schottky barrier process
and planar technology \cite{lin90}.

The PLUG calorimeter is instrumented with a total number of 8 planes
(16 modules), each containing 84 detectors and thus consists of
ideally 672 detectors (see table \ref{tabInst}).  For maintaining
simplicity and cost effectiveness it was not feasible to use a read
out system with this large number of individual electronic channels.
Instead two detectors situated in neighbouring planes along the beam
axis, but at the same radial position, are ganged together, thereby
reducing the number of electronic channels to 336. For similar
reasons, several detectors in one module (i.e. on the same read out
board) are connected to a single voltage supply channel. In total 80
such channels are used \footnote{CAEN modules (A336P, A436A) are
  used}.

\begin{table}[htbp]
  \begin{center}
    \leavevmode
    \begin{tabular}[t]{|c||c|c|c||c|}\hline
         &  S  & T  & R   &total \\ \hline\hline
      90 &  55 & -  & -   & 55 \\
      91 & 167 & 21 & 9   & 197 \\
      92 & 182 & 51 & 19  & 252 \\
      93 & 108 & -  & 36  & 144 \\ \hline
      sum & 512 & 72 & 64 & 648 \\ \hline
     \end{tabular}
  \end{center}
  \caption{Summary of installed detectors of the PLUG in 94 sorted by
           production year and detector shape: square (S), triangular (T) and
           rectangular (R).}
  \label{tabInst}
\end{table}

\newpage

\section{Basic Characteristics of the Used Silicon Detectors}
\label{section3}

The capacitance of a silicon detector with area A and a depleted
thickness X (depth of the electric field) is simply given by eq.\
(\ref{eq1}), where $\epsilon_{0}=8.854 \cdot 10^{-12} F m^{-1}$
and $\epsilon_{Si}=11.9$.
\begin{equation}
  \label{eq1}
  C = \frac{\epsilon_{Si} \epsilon_{0} A}{X}
\end{equation}
X is a function of the applied reverse bias voltage V according to
eq.\ (\ref{eq2})
\begin{equation}
  \label{eq2}
  X = \sqrt{\frac{2\epsilon_{Si}\epsilon_0}{e\left|N_{eff}\right|}V}
  \quad \mbox{,} \quad
  V \le V_{d}
\end{equation}
where e is the elementary charge and $N_{eff}$ the effective impurity
concentration of the base silicon material. For the depletion of the
total detector thickness d we get from eq.\ (\ref{eq2}) the necessary
bias voltage.
\begin{equation}
  \label{eq3}
  V_d = \frac{e\left|N_{eff}\right|}{2\epsilon_{Si}\epsilon_0} d^{2}
\end{equation}
The voltage dependance of the detector capacitance is hence given by
\begin{equation}
  \label{eq4}
  C (V) =
  \left\{
    \begin{array}{r@{\quad,\quad}l}
      A \sqrt{\frac{\epsilon_{Si}\epsilon_0 e\left|N_{eff}\right|}{2V}} & V <
V_{d}\\
      A \frac{\epsilon_{Si}\epsilon_0}{d} & V \ge V_{d}
    \end{array}
  \right.
\end{equation}
In this simple description for a Schottky barrier diode we have
neglected the "built in" voltage which should have to be added to the
external bias voltage V but is usually quite small (appr.~0.5 V)
compared to the depletion voltage (around 50 V for our detectors).

Due to the fact that the front electrode partially overlaps the oxide
passivated edge of the silicon detector, the C-V (capacitance/voltage)
characteristic given by eq.\ (\ref{eq4}), has to be modified for low voltages.
This overlapping "field plate" region forms a MOS (Metal-Oxide-Silicon)
capacitance in parallel to the diode itself.

For voltages smaller than the "flat band voltage" $V_{fb}$ we still
have an accumulation layer underneath the $SiO_{2}-Si$ interface and
due to the extremely small thickness of the silicon oxide passivation
layer (appr.~2000~\AA) the MOS-capacitance (to be added to the diode
capacitance of eq.~(\ref{eq4})) is very high.  The silicon bulk below
the field plate only becomes depleted once the external bias voltage
exceeds the flat band voltage~: the C-V characteristic is then
dominated by eq.\ (\ref{eq4}). Depending on the thickness and quality
of the oxide passivation typical flat band voltages range betweeen 1
and 5 V and are hence not problematic with respect to the much higher
operation voltages.

The other important performance parameter of the silicon detectors is
the reverse current $I$, which is in our case composed of three main
contributions.  Normally dominant is the bulk part, which is caused
by thermal electron--hole--pair generation throughout the depleted
volume of the detector. According to its thickness (eq.\ (\ref{eq2}))
we get
\begin{equation}
  \label{eq5}
  I = i (T) A
      \sqrt{\frac{2\epsilon_{Si}\epsilon_0}{e\left|N_{eff}\right|}V}
  \quad \mbox{,} \quad
  V \le V_{d}
\end{equation}
and for $V \ge V_d$ a constant value of $I = i(T) \cdot A \cdot d$.
$i(T)$ is the temperature dependant generation current density.  In
addition to the simple behaviour of eq.\ (\ref{eq5}) we have two
surface contributions. One is related to the $SiO_{2}$--Si interface
of the field plate region of the detector rim. Above the flat band
voltage the interface states located there, which also act as
generation centres for electron--hole--pairs, become accessible to the
electric field and hence contribute to the overall reverse current.
Normally this is seen as a certain jump in the I-V (current/voltage)
characteristic at the flat band voltage. The size of this effect is
proportional to the overlapping field plate region and depends also on
the quality of the oxide (concentration of interface states). The
second contribution to the reverse current is due to the Schottky
barrier itself. In contrast to the interface generation, this part
exhibits also a certain voltage dependence and the size of its
contribution is governed by the "barrier height" which is a measure of
the Schottky barrier quality.  For more details on both C-V and I-V
characteristics see \cite{lin90}.

\begin{figure}[htbp]
   \begin{center}
     \begin{picture}(150,300)
       \put(-150,35){\epsfig{file=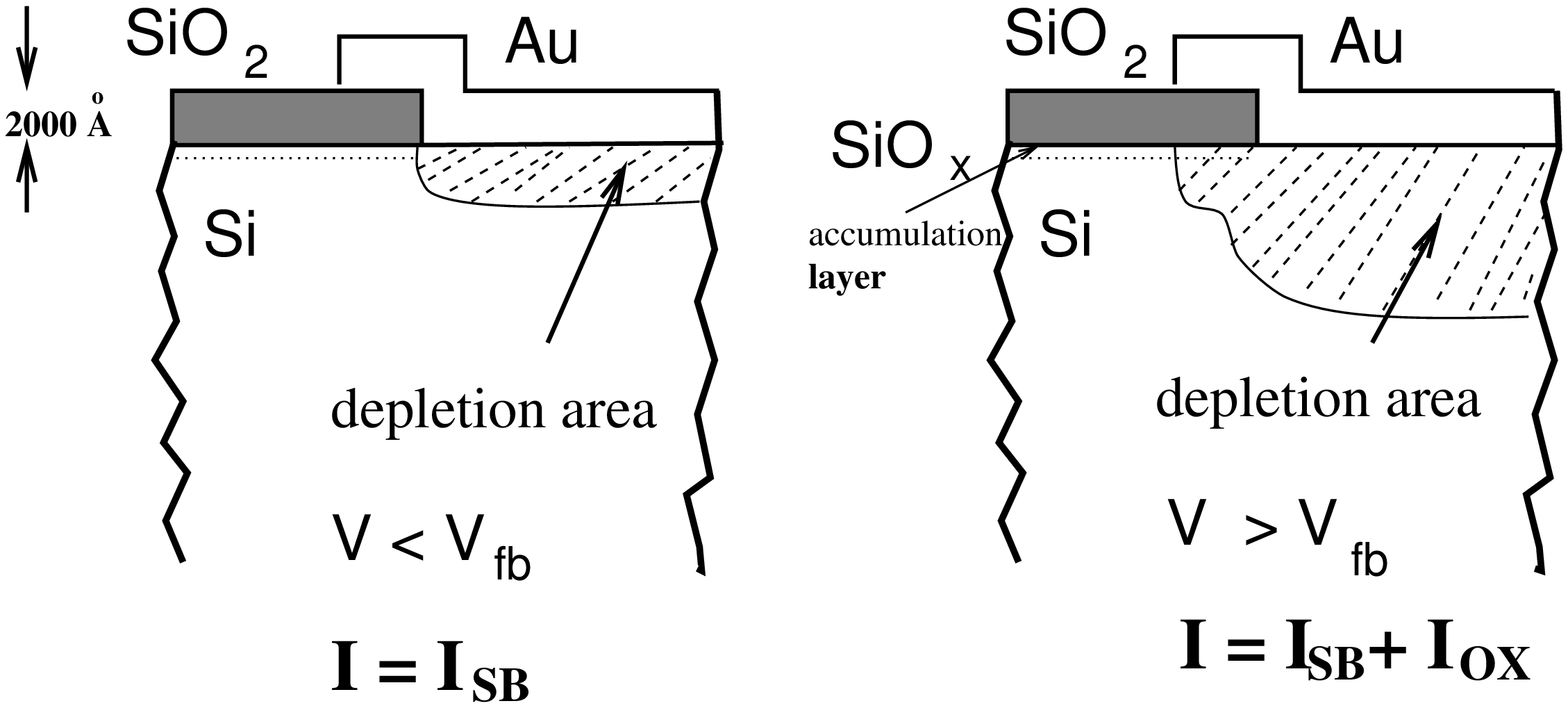,width=17cm,
                             height=18cm,angle=90}}
     \end{picture}
   \end{center}
   \caption[Schematic cross section of a detector]{Schematic cross
            section of one detector, the border of the depleted region is
           indicated for different voltages.}
   \label{detcross}
  \vspace{1cm}
  \begin{center}
    \epsfig{figure=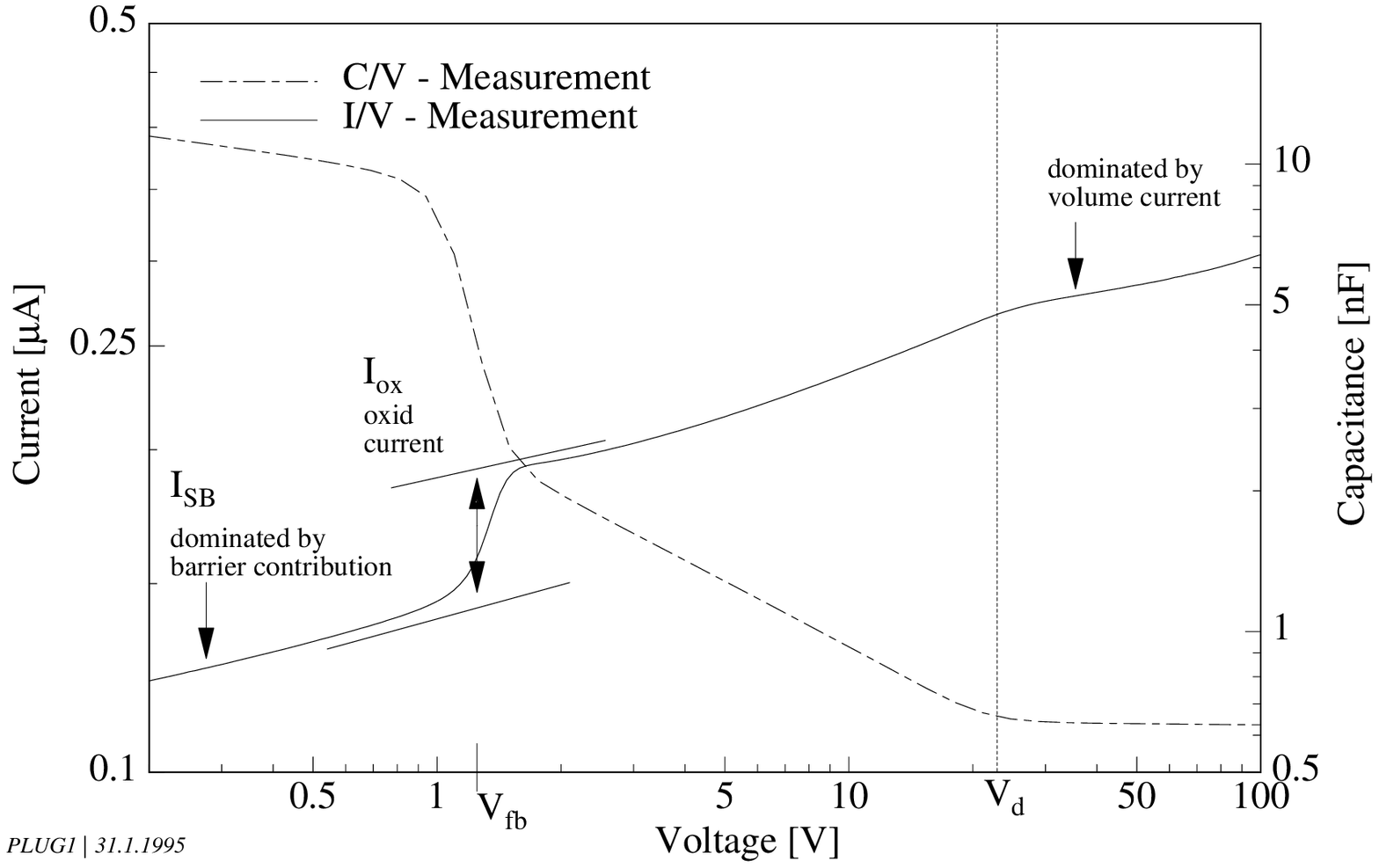,height=10cm,width=10cm}
  \end{center}
  \caption[Field and Characteristics]{The I-V and C-V characteristics
           are given for a typical PLUG detector.}
  \label{ivcvchara}
\end{figure}

In the figures \ref{detcross} and \ref{ivcvchara} the features discussed above
are shown for a typical PLUG detector after all technological process
steps have been completed but prior to being used during actual beam
runs. Figure \ref{detcross} visualizes the principal electric field
extension at the detector edge below and above flat band voltage, the
C-V and I-V characteristics are shown in fig. \ref{ivcvchara}. In the
given example the flat band voltage turns out to be 1.5 V. This is
both seen in the dramatic decrease of the detector capacitance and the
jump of the reverse current at that point. For larger voltages the
capacitance turns out to be inversely proportional to $\sqrt{V}$ as to
be expected by eq.\ (\ref{eq4}) and reaches its geometrical limit of
630 pF at the depletion voltage ($V_{d}=22 V$), while the current
saturates at the same point.

\section{PLUG Readout Circuit}
\label{PLUGRC}

The readout circuit of the PLUG calorimeter was designed in such a way
that a maximum of vital information about the individual silicon
detectors could be obtained during the beam period. Specifically the
voltage needed to extend the electric field throughout the detector
thickness and hence to ensure the total collection of charge generated
by a penetrating particle, should be checkable at any time. This
depletion voltage can however not be measured as in the laboratory by
using e.g. a capacitance bridge.  Instead the response of the readout
circuit to a given charge injected by a testpulser is used. Details
are given below.

Another similarly important parameter is the reverse current of each
silicon detector. Only if this value is known one can calculate the
voltage drop across the bias resistor and hence ensure that the net
detector bias voltage stays sufficiently above the depletion voltage.
As two detectors situated in successive planes (modules), but at the
same position are ganged together for readout and up to 14 detectors
in one single plane are connected to one bias channel (see section 1),
the monitoring of both the C-V and I-V characteristics for individual
detectors imposes a significant difficulty, which finally led to the
following compromise solution.

\begin{figure}[htb]
  \begin{center}
  \epsfig{figure=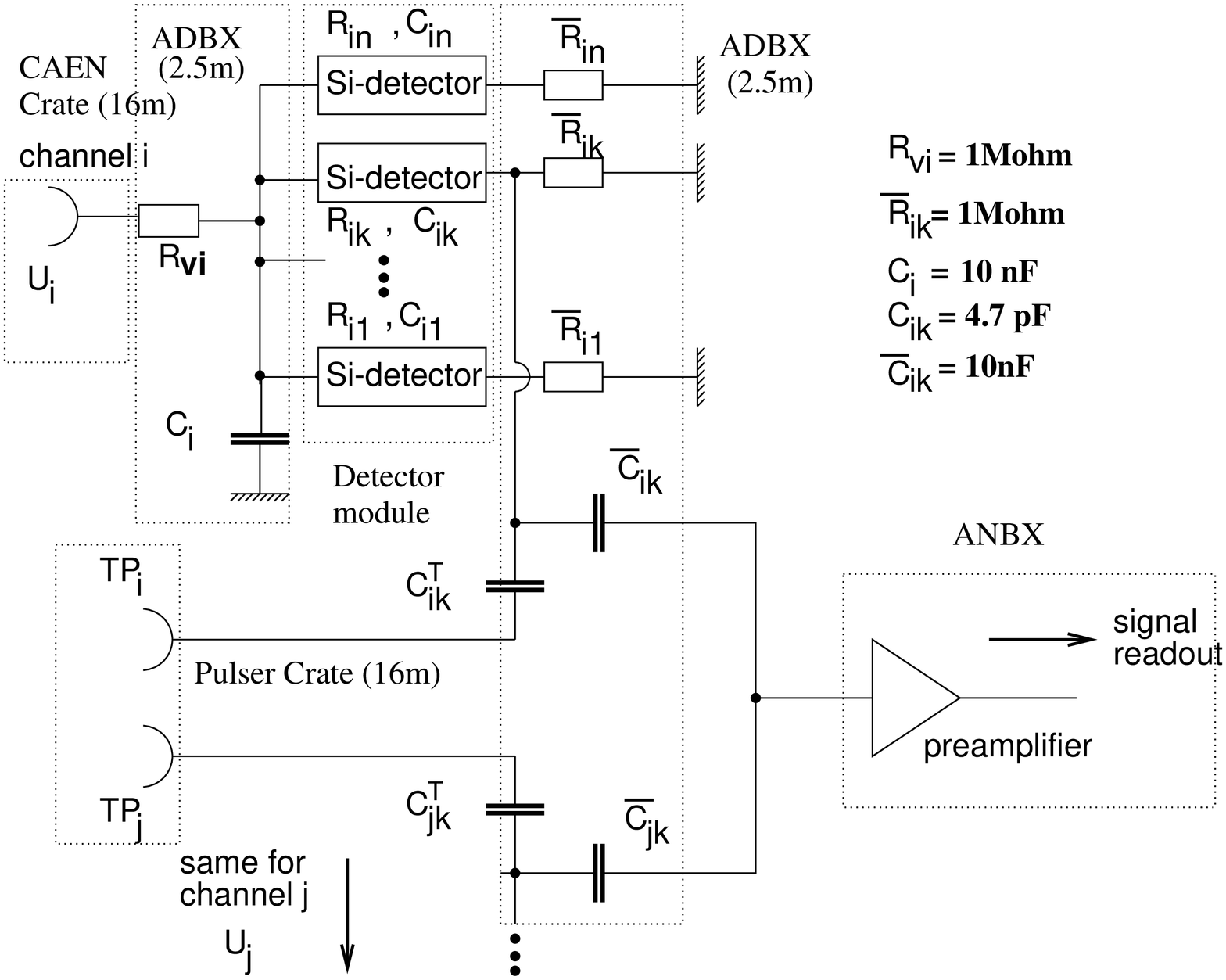,height=9cm,angle=90}
    \vspace{1cm}
    \caption[PLUG circuit diagram]
      {Simplified circuit diagram of the PLUG readout for a given bias
       channel i and signal readout k.}
  \label{fig5}
  \end{center}
\end{figure}

In fig.\ \ref{fig5} a simplified circuit diagram of one read out
channel is shown. The signal output of two detectors located in
adjacent planes, for instance in the first and second, but at the same
r, $\Phi$ position of the calorimeter, are connected to one
preamplifier. An appreciable ganging had also to be used for the bias
supply of all 672 detectors from only 80 available bias channels.
Unfortunately a bias distribution box, as used for the LAr calorimeter
had to be ruled out because of unacceptable limitations for the
detector current. Due to the exposed position of the PLUG calorimeter
in the very forward angular range, some radiation damage had to be
expected with resulting larger increase of the reverse current for the
innermost detectors closest to the beam pipe. In order to achieve a
similar accuracy in current monitoring throughout the calorimeter, a
bias ganging of only 3 detectors was foreseen for the innermost
detectors, extending to a connection of up to 14 detectors to one bias
supply at the outer rim.  At a given operational voltage each detector
may be symbolized by its capacitance $C_{ik}$ and the reverse current
by a parallel resistor $R_{ik}$.  A common filter network
$R_{vi},C_{i}$ for the $n$ detectors, belonging to one bias module, is
used for the reduction of low frequency noise picked up via the long
bias transfer line between the unit located in the electronic trailer
of the H1 experiment and the PLUG. The individual bias resstors
${\overline R_{ik}}$ on the other side of the detectors enable the
signal readout via the likewise individual coupling capacitors
${\overline C_{ik}}$. For calibration purposes a charge is induced via
the charge terminator capacitance $C_{ik}^{T}$. It has to be noticed,
that the actual realisation of the block diagram is contained in 5
different electronic units (indicated by dotted lines in
fig.~\ref{fig5}) connected by cables of varying length. The bias unit
and the pulser crate are located in the electronic trailer, the
adapter box ADBX (bias distribution and readout ganging) and the
analog box ANBX sit behind the return yoke of the magnet and the PLUG
calorimeter itself.

Especially the cables between the detectors and the coupling capacitors
together with any inherent stray capacitance add a constant value of
$C_{S,ik} \approx 300 \mbox{pF}$ to the voltage dependent detector capacitance
$C_{D,ik}$.
Thus $C_{ik}$ in fig.\ \ref{fig5} is the sum of both these values.
\begin{equation}
  \label{eq6}
  C_{ik} = C_{D,ik} + C_{S,ik}
\end{equation}

\section{Monitoring of Diode Characteristics}
\label{monit}

In this section we will descibe the possibilities for monitoring the
current, capacitance and noise characteristics of the installed
detectors even during luminosity periods. The principal difficulty here
is that due to the necessary ganging described above, the individual
detector performance cannot be measured. The following discussion is given
for one bias channel i only; its index is omitted.

\subsection{Reverse current}
As $n$ detectors (n=3-14, see section \ref{PLUGRC}) are connected
to one bias supply unit, we cannot measure the individual detector
currents but have instead to assume that the measured current
$I$ is equally distributed, resulting in~:
\begin{equation}
  I_{k} = \frac{1}{n} I
  \label{eq7}
\end{equation}
Hence the net bias voltage for one detector is estimated to be
\begin{equation}
 V_{k} = V - \frac{1}{n} \, I \, {\overline R_{k}} -
 I \, R_{v}
  \label{eq8}
\end{equation}
Equations (\ref{eq7}) and (\ref{eq8}) enable us to measure the
I-V characteristics for each detector. We add here that the
correction due to the voltage drop across the common resistor $R_{v}$
is exact. However the term due to the individual preamplifier bias
resistor ${\overline R_{k}}$ is sensitive to the assumption of eq.(\ref{eq7})
and may be particularly important if the measured current is very
large.

\subsection{Detector capacitance}
\begin{figure}[hbt]
  \begin{center}
    \vspace{1cm}
    \begin{picture}(50,100)
      \put(50,-150){\epsfig{file=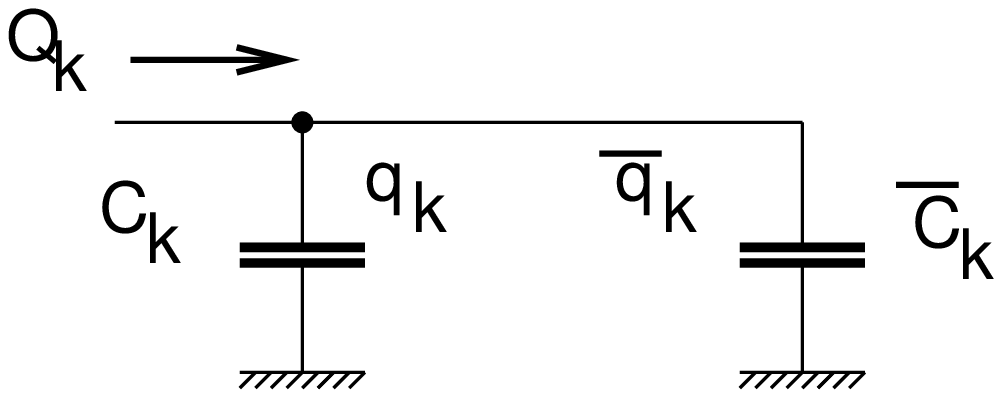,height=3cm,angle=90}}
    \end{picture}
    \vspace{-1cm}
  \end{center}
  \caption{Equivalent circuit diagram for the measurement of the
           detector capacitance.}
  \label{fig6}
\end{figure}

The C-V characteristic for each individual detector can be investigated
by measuring the signal height for a given input charge as function
of the detector voltage. A charge $Q_{k}$ is induced by the test pulser
via the charge terminator $C_{k}^{T}$. Taking the values of the
different capacitances into account (fig.~\ref{fig5}), the response of the
preamplifier can be calculated according to the equivalent circuit
diagram of fig.~\ref{fig6}. This results in the following relation
 \begin{equation}
   \frac{q_{k}}{C_{k}}=\frac {\overline q_{k}}{\overline C_{k}}=
   \frac{Q_{k}}{C_{k}+{\overline{C_{k}}}}
   \label{eq9}
 \end{equation}
where ${\overline q_{k}}$ is the input charge for the preamplifier.
Thus the output signal $S_{k}$ is given by eq.(\ref{eq10}), where
$C_{k}$ is replaced by the expression of eq.(\ref{eq6}).
 \begin{equation}
   S^{-1}_{k} \propto \left( \overline q_{k} \right)^{-1} =
     \frac{C_{D,k} + C_{S,k} + {\overline C_{k}} }
          {\overline C_{k} \, Q_{k}}
   \label{eq10}
 \end{equation}
Using eq.(\ref{eq4}) and the discussion in section 2 the voltage
dependence of the signal can be finally represented by~:
\begin{equation}
  \label{eq11}
  S_{k}^{-1} =
  \left\{
    \begin{array}{r@{\quad,\quad}l}
      \approx C_{MOS} + b & V_{k} < V_{fb} \\
      a \, V_{k}^{-1/2} + b & V_{fb} <  V_{k} < V_{d,k} \\
      a \, V_{d,k}^{-1/2} + b & V_{k} \ge V_{d,k}
    \end{array}
  \right.
\end{equation}
It should be mentioned that the same uncertainties for the voltage
corrections as discussed for the I-V characteristic according to
eq.(\ref{eq8}) are relevant here too.

\subsection{Signal noise}

The signal noise can only be measured for each electronic channel,
connected to two detectors ganged together as seen in fig.~\ref{fig5}.
Capacitance and noise characteristics are measured simultaneously.
For the capacitance we use the mean value of the signal distribution
and the noise is extracted from its standard deviation. The noise
performance contains two contributions, one of which is due to the
input capacitance of the preamplifier and the second corresponds to
the detector noise. For newly commissioned detectors the second
contribution is negligible. Hence the combination of I-V (ganging in
one plane) and noise-characteristics (ganging of two detectors in two
consecutive planes) provides a tool to identify single bad detectors.

\newpage

\section{Measurements during HERA operation}
\label{heram}

The measurements taken during the 1994 HERA operation period
are presented. The section is subdivided into the results
concerning the reverse current, the detector capacitance,
the signal noise and the absorbed radiation dose.
These parameters have been monitored for each of the
648 detectors during the whole '94 run-period.

\subsection{Reverse Current}

\begin{figure}[htbp]
  \begin{center}
    \leavevmode
      \epsfig{figure=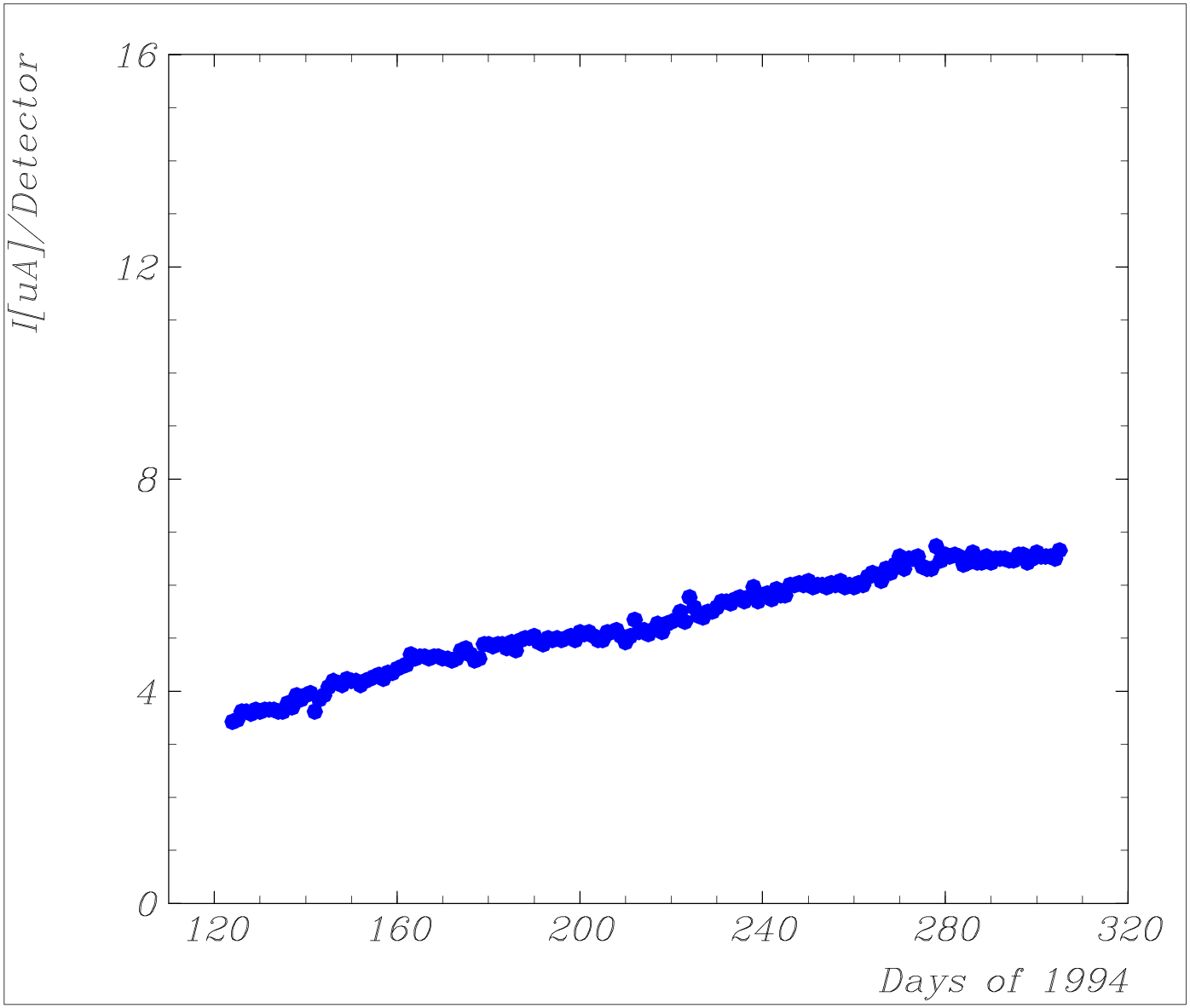,width=8cm,angle=90}
      \epsfig{figure=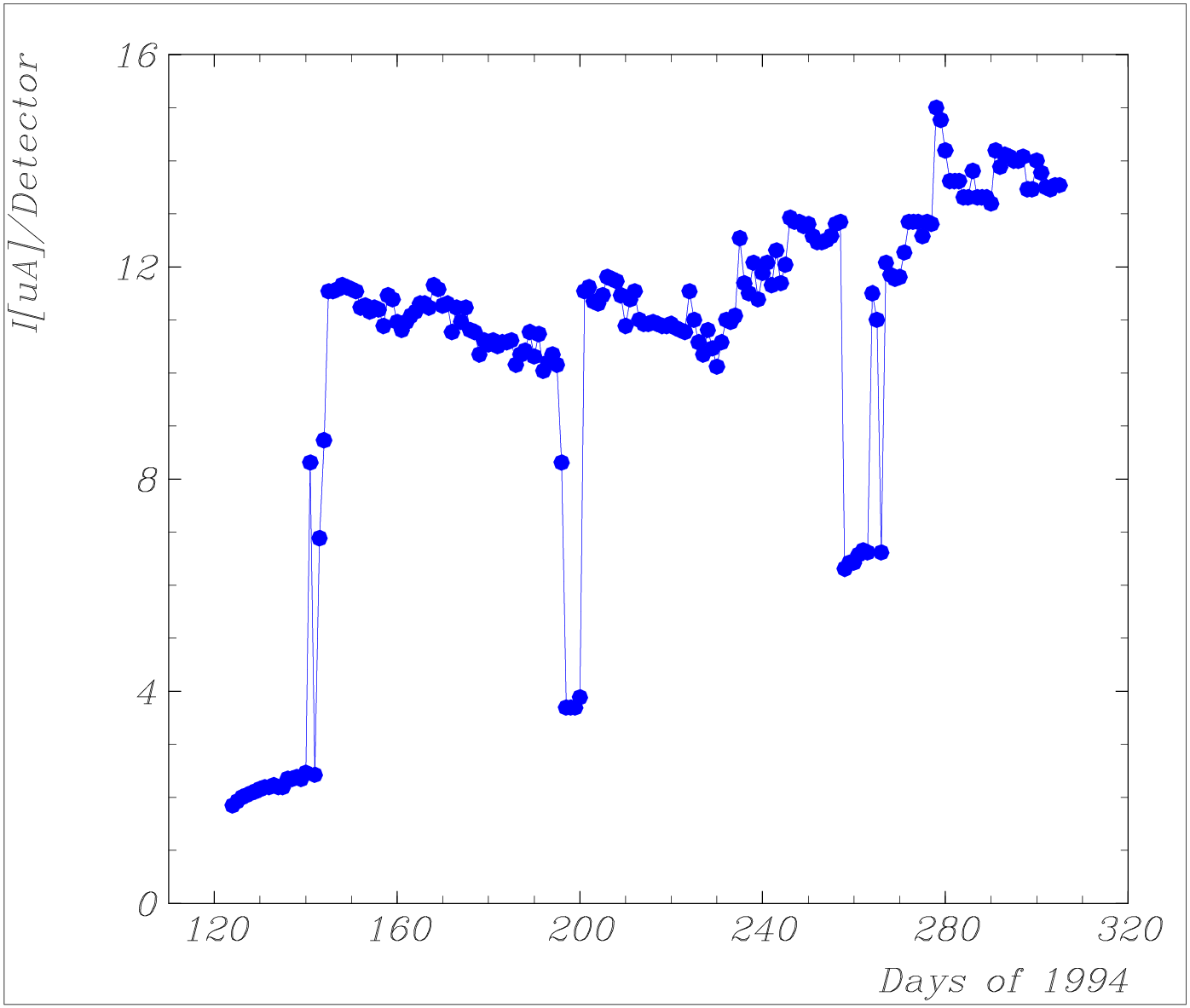,width=8cm,angle=90}
  \end{center}
  \caption[Reverse current, every day]{%
    Reverse current on a long time scale given for a normal ''good''
    bias-voltage-channel (left side) and one containing at least one ''bad''
    detector (right side), see text.}
  \label{figDark1}
  \begin{center}
    \leavevmode
      \epsfig{figure=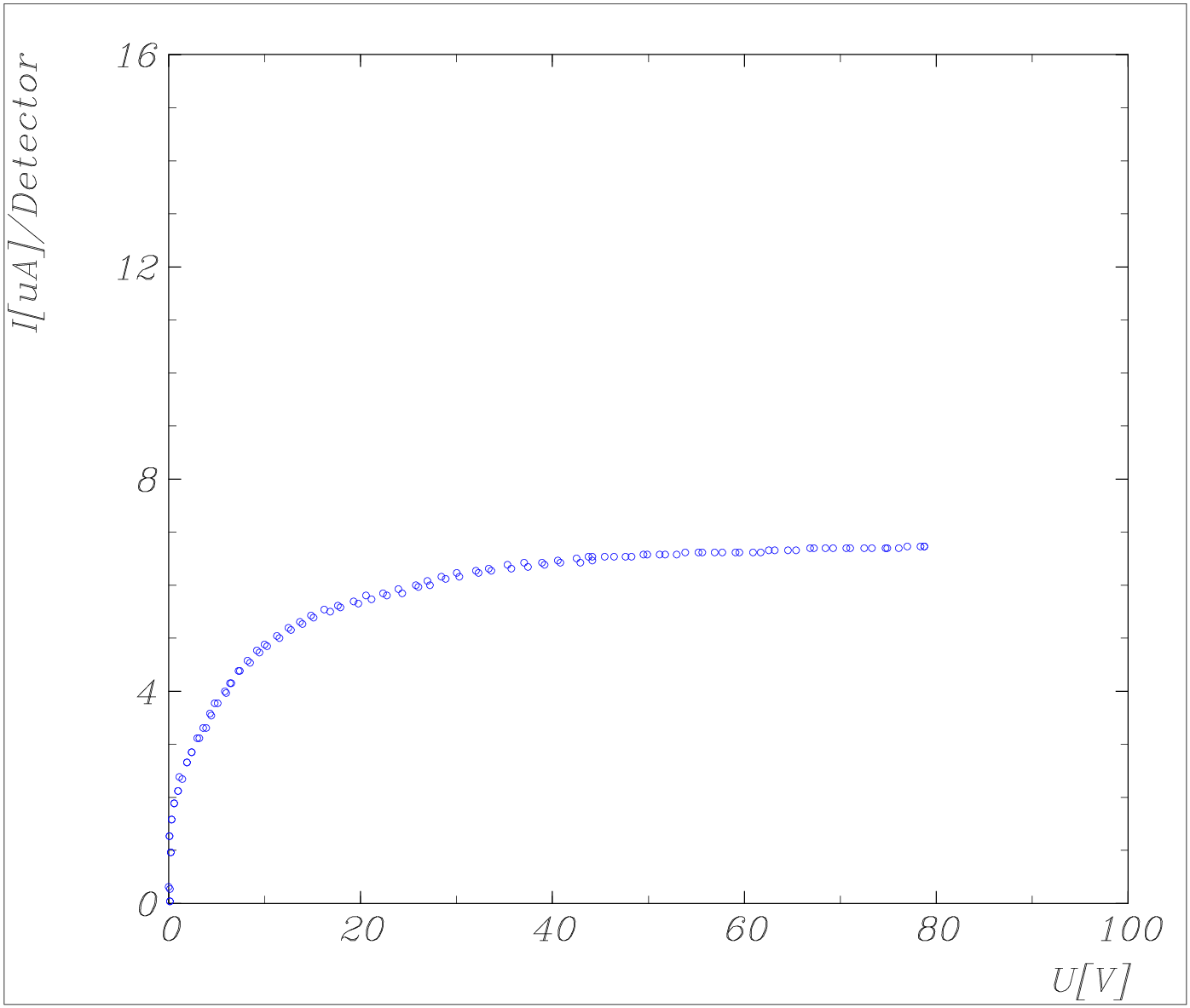,width=8cm,angle=90}
      \epsfig{figure=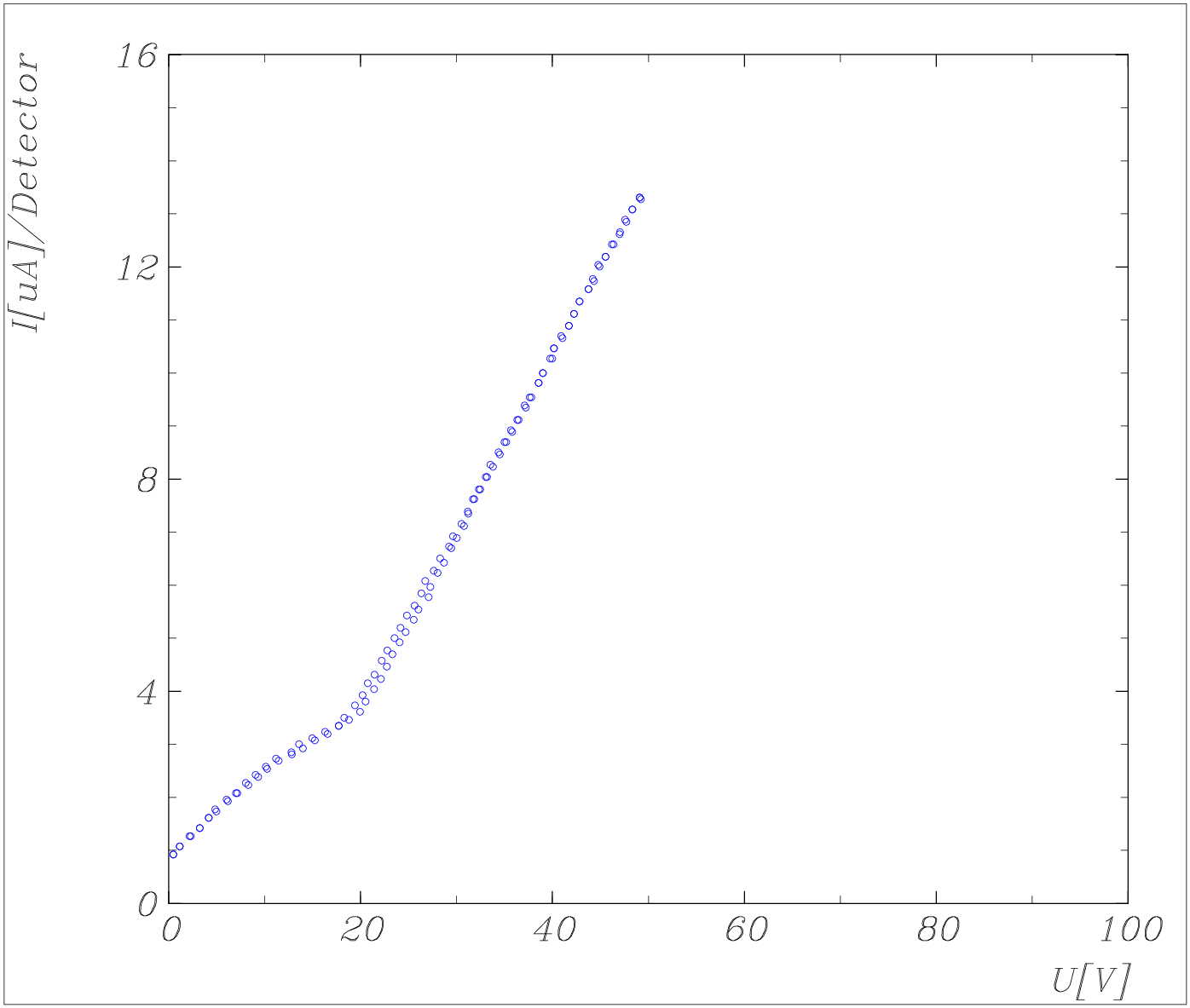,width=8cm,angle=90}
  \end{center}
  \caption{IV characteristics of a normal ''good'' bias-voltage-channel
           (left side) and one containing at least one ''bad'' detector
           (right side) showing break-down.}
  \label{selected_iv_curves}
  \begin{center}
    \leavevmode
      \epsfig{figure=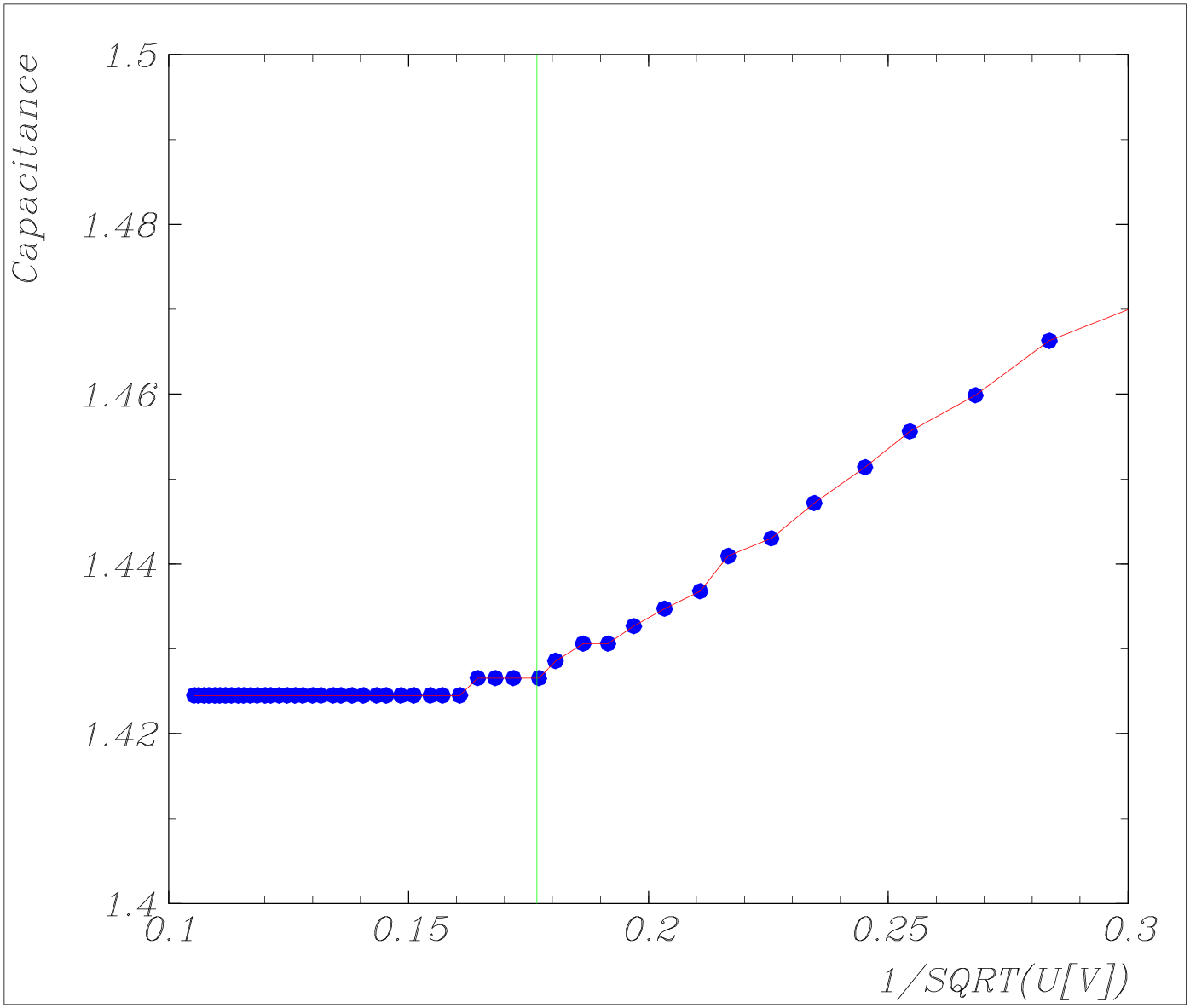,width=8cm,angle=90}
      \epsfig{figure=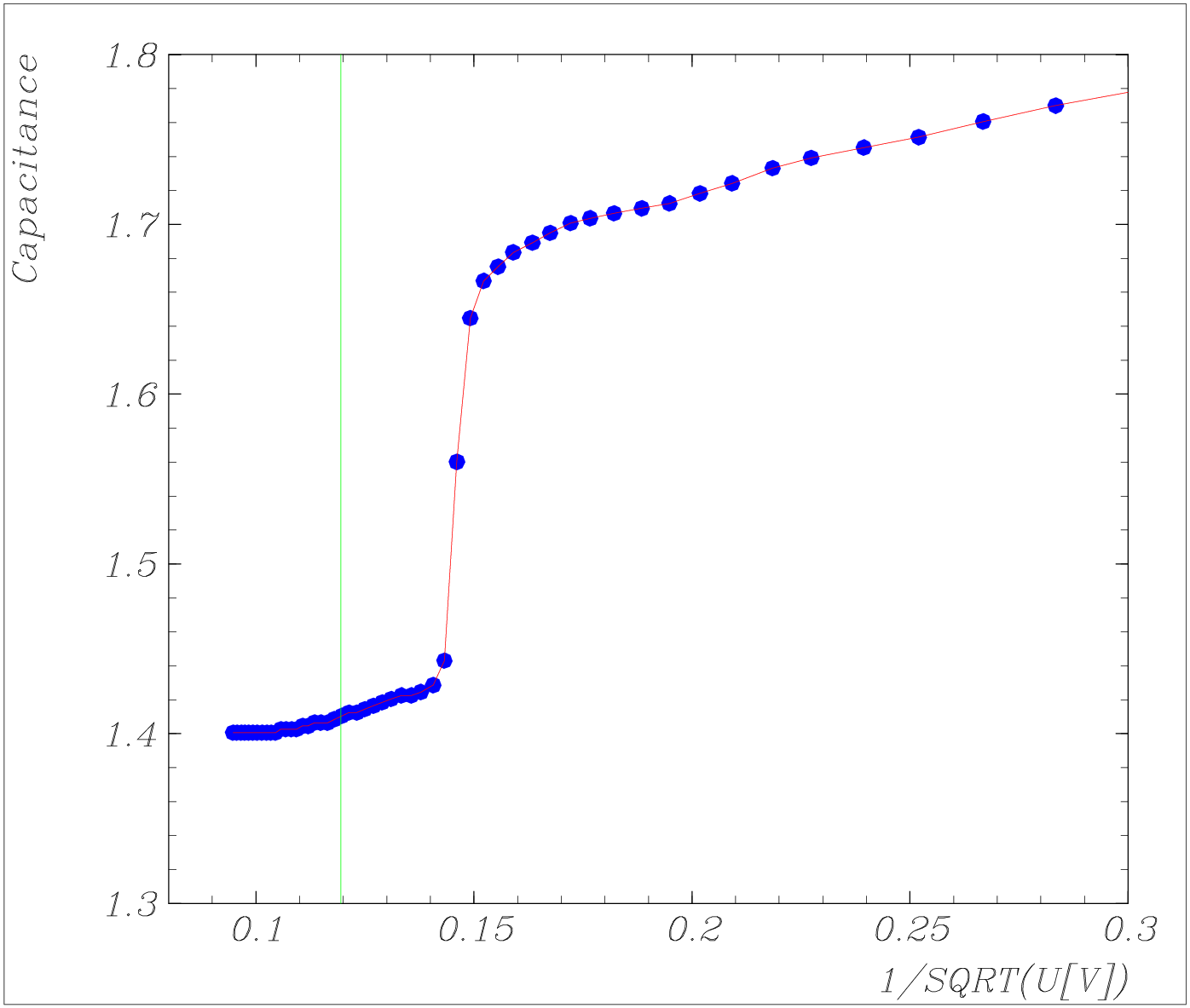,width=8cm,angle=90}
  \end{center}
  \caption[C-V characteristics as function of $1/\protect{\sqrt{V}}$]{%
    The normal C-V characteristic and the C-V characteristic for a
    detector with an increased flat band voltage are given as a
    function of $1/\sqrt{V}$. The capacitance is given in arbitrary
    units.}
  \label{figcvsqrt}
\end{figure}

In figure \ref{figDark1} the reverse current as a function of time
given in days is plotted for two typical bias-voltage-channels.  The
first channel shows a continuous development of the reverse current
with an almost linear tendency.  The increase of the reverse current
is due to an increase of the surface contribution as investigated e.g.
in ref. \cite{schulz}.  Synchrotron radiation creates new generation
centers of charges inside the interface (Si-$SiO_{2}$) causing a
higher surface current (see fig. \ref{detcross}).  This increase is
proportional to the absorbed dose.  Therefore a radial dependence of
the increase is observed with inner detectors showing the strongest
effect. Details are discussed in sect.~\ref{labom}.  The data were
taken on day 242 of the run period (compare fig. \ref{figDark1}).

The second channel shows a faster linear increase of the reverse
current and its evolution is distorted by strong jumps of the current.
These jumps are abnormal, they can be explained by a local break-down
of the depletion towards the border of a detector causing a short
circuit in the corresponding bias-voltage-ring. The jumps provide high
noise spoiling the measurement. Therefore detectors showing break-down
have to be removed from the readout \footnote{This is done by an
  automated procedure, "the hot channel monitor", which tracks the
  performance of each detector with time (see section \ref{noise})}.

In figure \ref{selected_iv_curves} the I-V characteristics of the same
two bias-voltage-channels are shown.  The reverse current of the first
channel has the typical dependence of the applied voltage. The curve
shows a decreasing slope until the depletion voltage is reached (fig.~
\ref{ivcvchara}) proving that all detectors connected to this channel
are operational.

The second channel shows the normal evolution up to values of about
20~V and above this voltage the I-V characteristic is governed by an
ohmic behaviour. No saturation of the current is observed indicating
that at least one detector of this channel shows a break-down .

\subsection{Detector Capacitance}
\label{cv_results}

Figures \ref{figcvsqrt} and \ref{figcvlin} illustrate the C-V
characteristics of two typical detectors. Figure \ref{figcvsqrt} shows
the C-V characteristic of a detector as a function of $1/\sqrt{V}$ and
figure \ref{figcvlin} the C-V characteristic as a function of the
voltage V.

The left part of fig. \ref{figcvsqrt} illustrates nicely the
saturation of the capacitance at depletion voltage as expected from
eq.(\ref{eq11}).  The depletion voltage of the shown detector is
measured to be $32 \pm 1$ V corresponding exactly to the value, which
was determined before installation in the laboratory.  The unchanged
depletion voltage proves that the bulk of silicon is not damaged. The
left part of fig. \ref{figcvlin} allows to study the flat band voltage
of the same detector. It is found to be 2~V. Also no change compared
to the measurement made in the laboratory is observed.

The right part of the figures \ref{figcvsqrt} and \ref{figcvlin}
represent a different detector with an increased flatband voltage of
$45 \pm 2$ V. This increase is due to radiation damage of the
$SiO_{2}$ passivation at the border of the detector. The observed
flatband voltage is lower than the depletion voltage of 82 V, still
visible in the C-V characteristic.

The increase of the flatband voltage induced by ionising radiation is
a known effect. In the particular case of the PLUG detectors the
increase has been reproduced by independent measurements
\cite{pschli95}. The increase has not been observed for all detectors.
This will be discussed in section~\ref{discu}.

\subsection{Signal Noise}
\label{noise}

\begin{figure}[htbp]
  \begin{center}
    \leavevmode
      \epsfig{figure=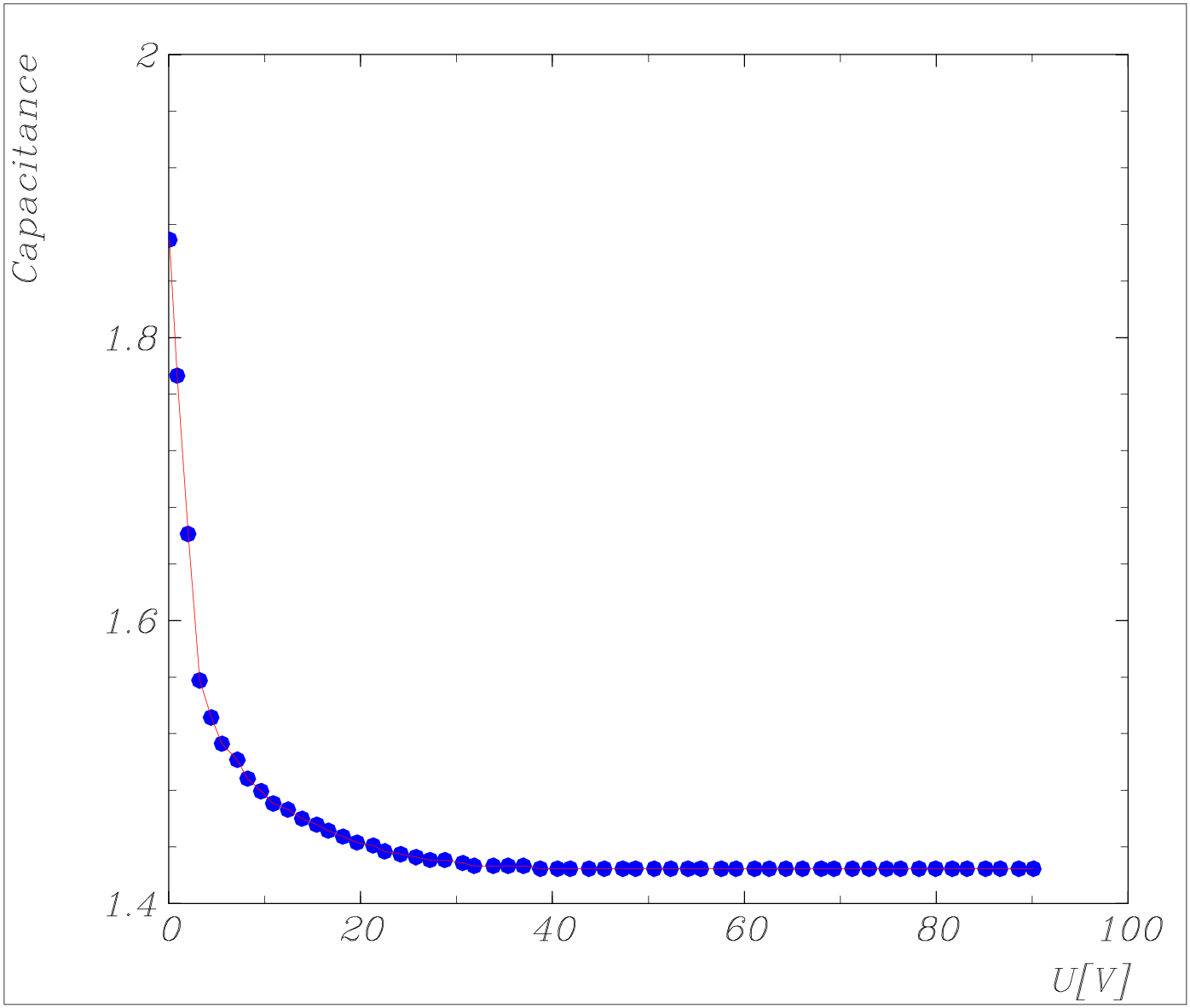,width=8cm,angle=90}
      \epsfig{figure=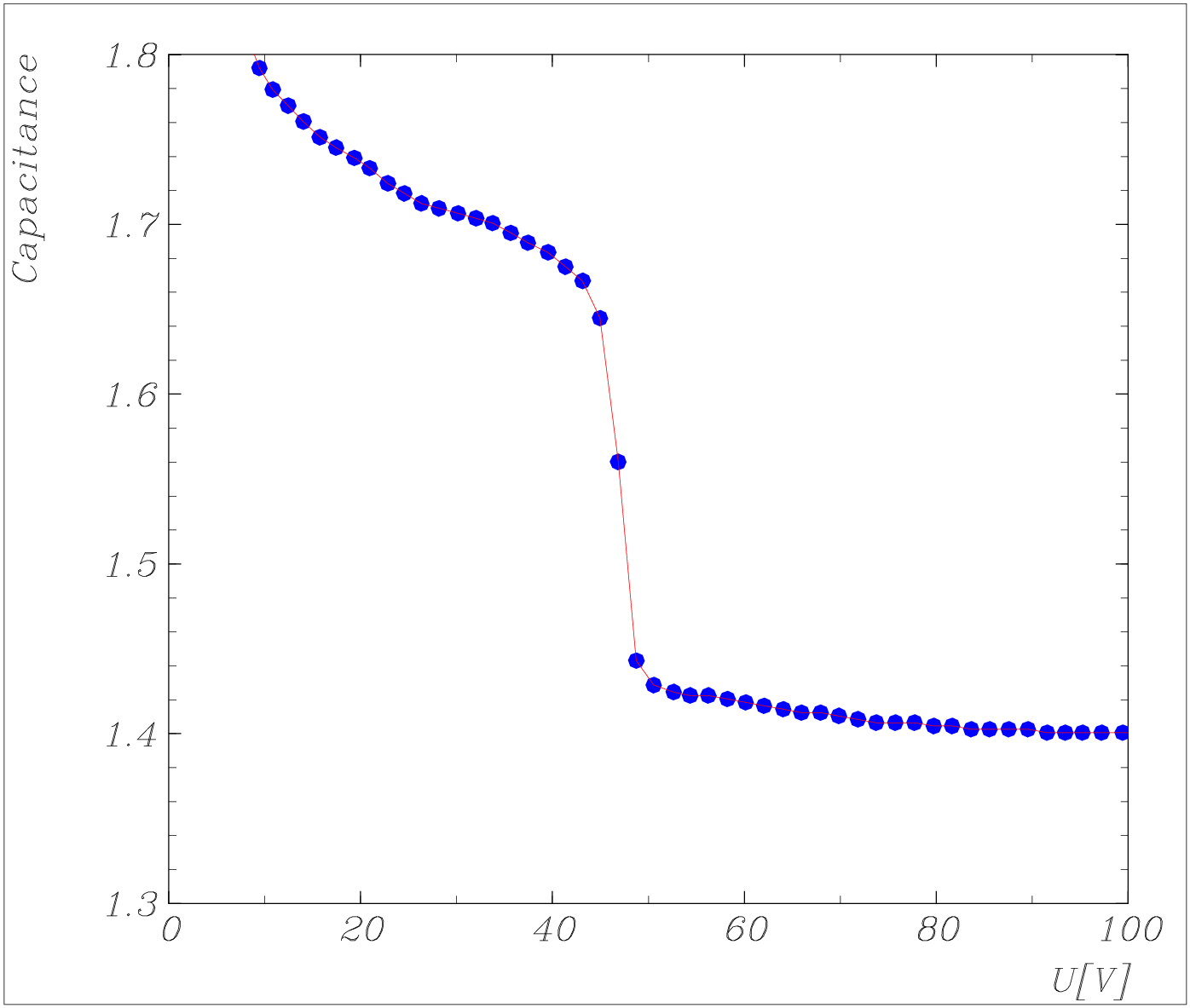,width=8cm,angle=90}
  \end{center}
  \caption[C-V characteristics as function of the voltage $V$]{%
    The normal C-V characteristic and the C-V characteristic for a
    detector with an increased flat band voltage are given
    as a function of the voltage $V$.}
  \label{figcvlin}
  \begin{center}
    \leavevmode
      \epsfig{figure=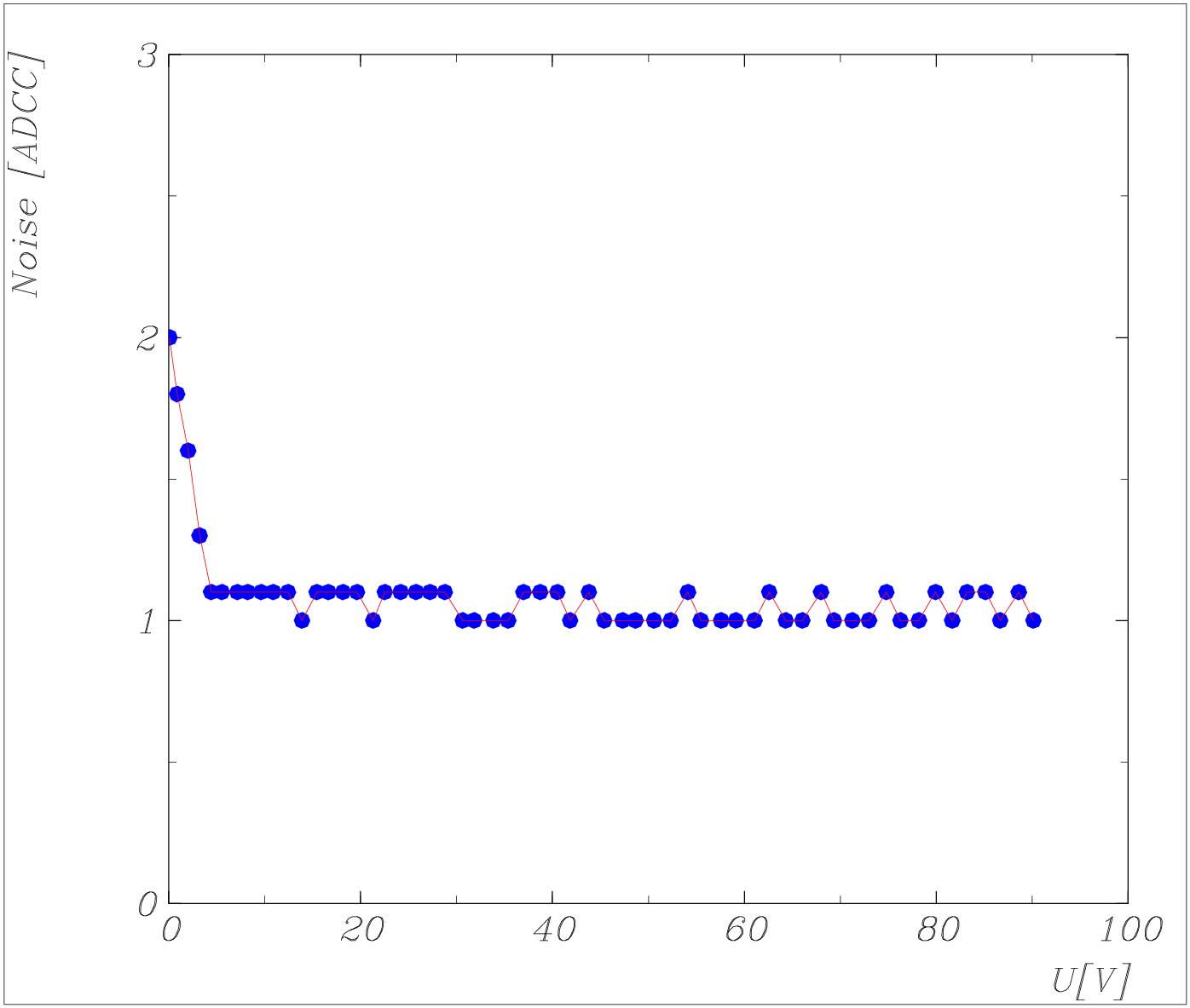,width=8cm,angle=90}
      \epsfig{figure=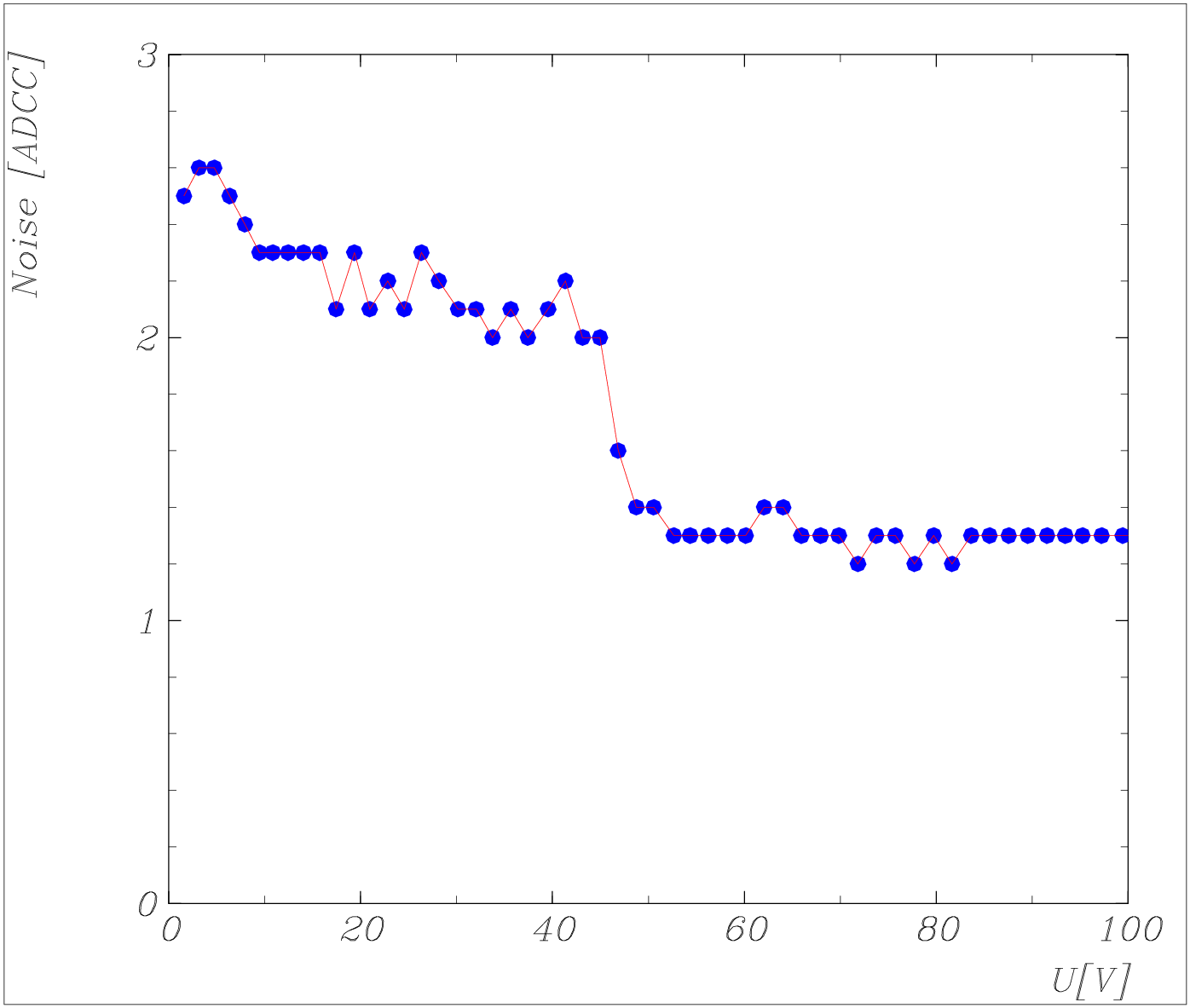,width=8cm,angle=90}
  \end{center}
  \caption[The detector noise as a function of the voltage $V$]{%
    The noise as a function of the voltage $V$ is given
    in units of ADC counts for both preceding detectors.}
  \label{fignoise1}
  \begin{center}
    \leavevmode
    \epsfig{figure=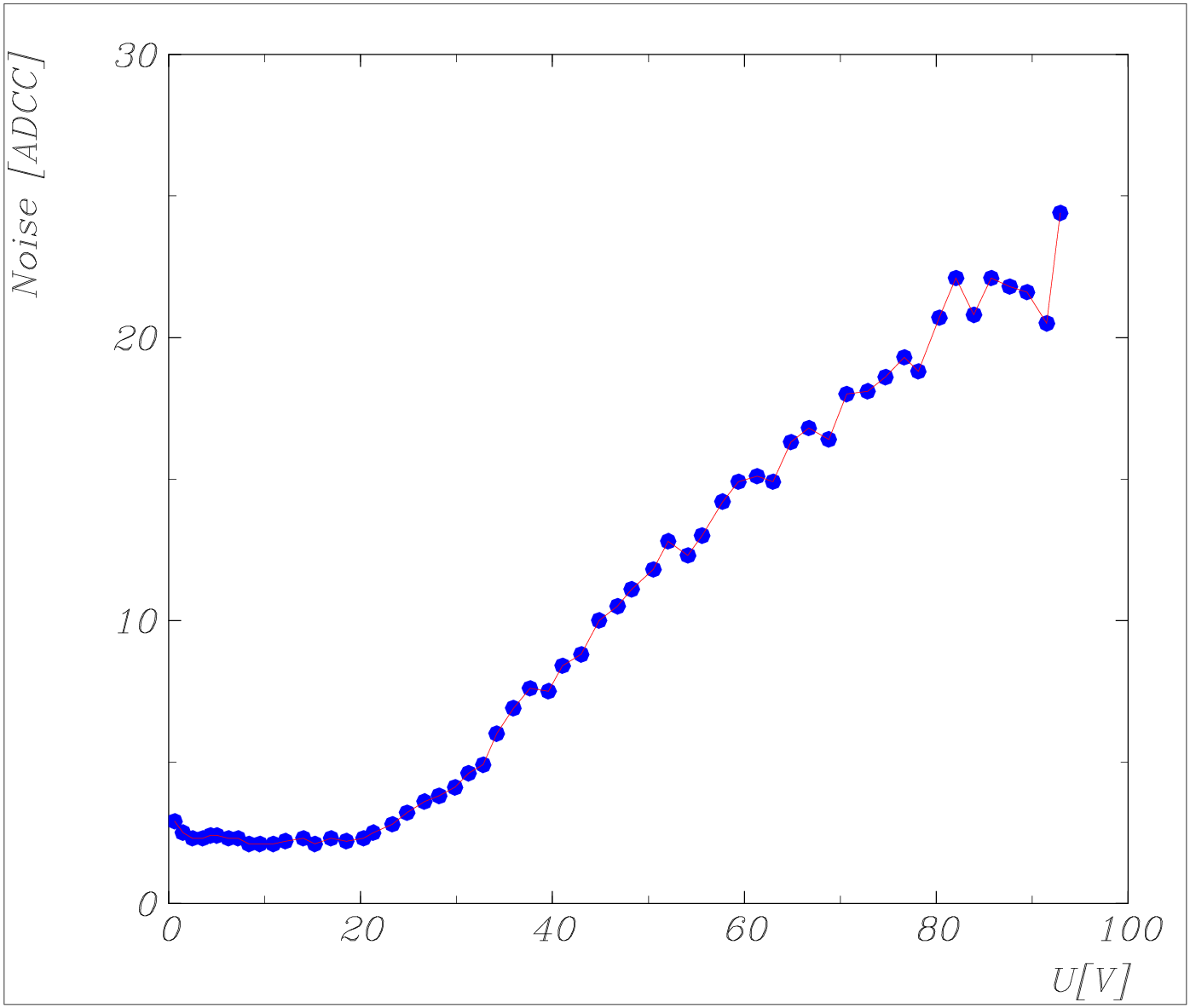,width=8cm,angle=90}
  \end{center}
  \caption[The detector noise of a broken-down detector]{%
    The noise as a function of the voltage $V$ is given
    in units of ADC counts for a detector showing break-down.}
  \label{fignoise2}
\end{figure}

The detector noise is shown in figure \ref{fignoise1} for both of the
preceding detectors as a function of the voltage $V$.  It follows for
both the evolution of their C-V characteristics (see fig.
\ref{figcvlin}). This is purely an effect of the used preamplifiers
being sensitive to the detector capacitance, so that the detector
noise itself can be neglected for undamaged detectors.  The first
detector has a noise of one ADC count corresponding to $\sim 8300$
electrons. The noise of the second detector drops down at the voltage
which corresponds to the increased flat band voltage and it also
stays constant above this value.

Figure \ref{fignoise2} shows the noise performance of a detector,
which is not operational due to high noise. Above 20 V indicating
break-down the noise increases linearly up to 30 ADC counts ($\sim$
1GeV).  This observation is in good agreement with the break-down of
the I-V characteristic in figure \ref{selected_iv_curves}.  This
detector has of course to be excluded from the readout in order not to
spoil the energy measurement.

\subsection{Absorbed Radiation Dose}
\label{dose}

The intention was to measure the integrated radiation dose produced by
ionizing particles at different radial positions. Therefore several
thermoluminescence dosimeters were installed inside the seventh slot
of the PLUG calorimeter during the 1994 run period.  The measured dose
varied from 300 Gy for the inner detectors to 10 Gy for the outer ones
\footnote{During the same time the total integrated luminosity for '94
  of 4 $pb^{-1}$ was collected.}.  This measured dose is not high
enough to explain the observed radiation damage.  Such effects were
expected at much higher dose values in the order of several kGy's.  It
is therefore assumed that the used dosimeters are not sensitive to the
radiation contributing most to the dose absorbed by the detectors.
\footnote{The used dosimeters RPL glastype R1T are only sensitive down
  to 30 keV.}. Instead the major effect is most likely due to very low
energetic electrons and photon from synchrotron radiation. This effect
is currently being investigated in more detail \cite{mbuck95}.

\section{Measurements in the Laboratory}
\label{labom}

All discussed measurements have been repeated in the laboratory after
the end of the run period. This had the advantage that every single
detector can be individually investigated and that no correction is
needed due to the ganging of several detectors to one bias supply unit
(sect. \ref{monit}). The reverse current, the detector capacitance as
well as the signal noise have been studied once again for each
detector. The results of this final analysis of the '94 run period are
presented in the following.

\subsection{Reverse Current}

The I-V characteristic of the 648 detectors installed in the PLUG
during the 94 run period has been analysed. $7~\%$ of the detectors
show a break-down as demonstrated in fig. \ref{selected_iv_curves} and are
no longer operational anymore.

\begin{figure}[htbp]
  \begin{center}
    \leavevmode
      \epsfig{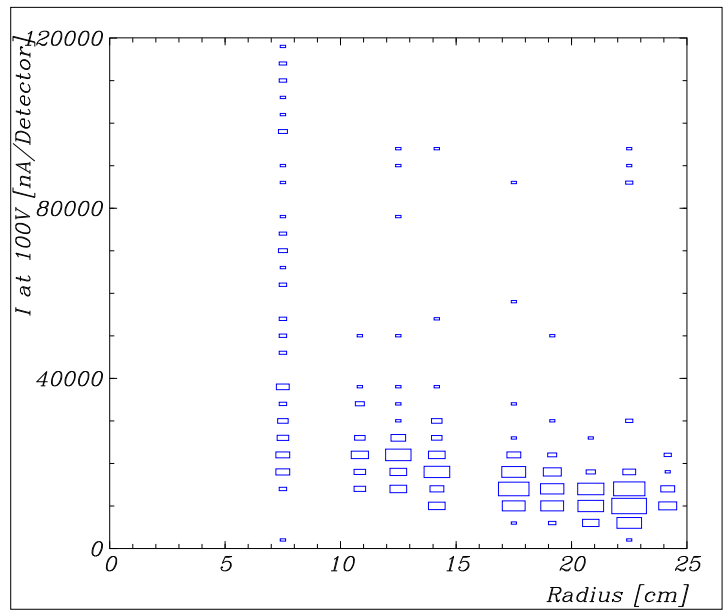}
  \end{center}
  \caption[Correlation of reverse current and radial position]{%
    The detector reverse current at 100 V is correlated with
    the radial position of the detector.}
  \label{radiv100}
\end{figure}

In figure \ref{radiv100} the correlation of the reverse current
measured at 100 V and the radial position is shown for the remaining
$92~\%$ of detectors. A clear radial dependence of the increase of the
reverse current is observed. The closer a detector is mounted to the
beampipe the higher is its current. $14~\%$ of the detectors have a
current higher than 6 $\mu A$. Most of these detectors are situated in
the innermost ring.  In addition to the radial dependence it should be
mentioned that detectors in the most backward plane where the electron
beam hits first are damaged more than detectors of the front plane,
through which the incoming proton beam passes first.  This indicates
once more that the increase of the reverse current is caused by
electromagnetic radiation.

\subsection{Detector Capacitance}

The C-V characteristics of all detectors have been studied in the
laboratory.  No increase of the depletion voltage has been observed.

\begin{figure}[htbp]
  \begin{center}
    \leavevmode \epsfig{figure=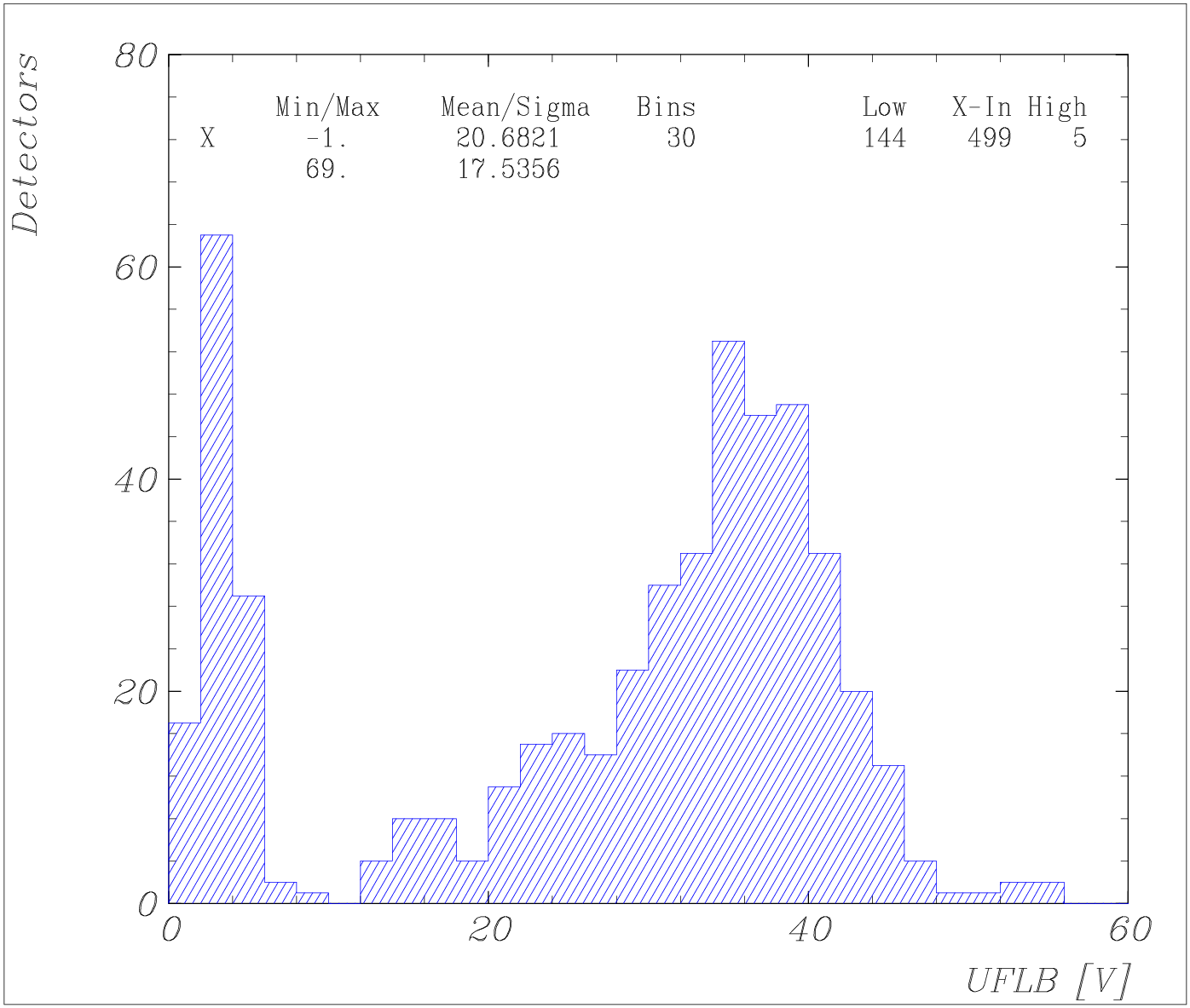,width=8cm,angle=90}
  \end{center}
  \caption[Distribution of flat band voltages]{%
    The distribution of the flat band voltages at the end of the
    '94 run period is presented.}
  \label{figflbvo}
  \begin{center}
    \leavevmode
      \epsfig{figure=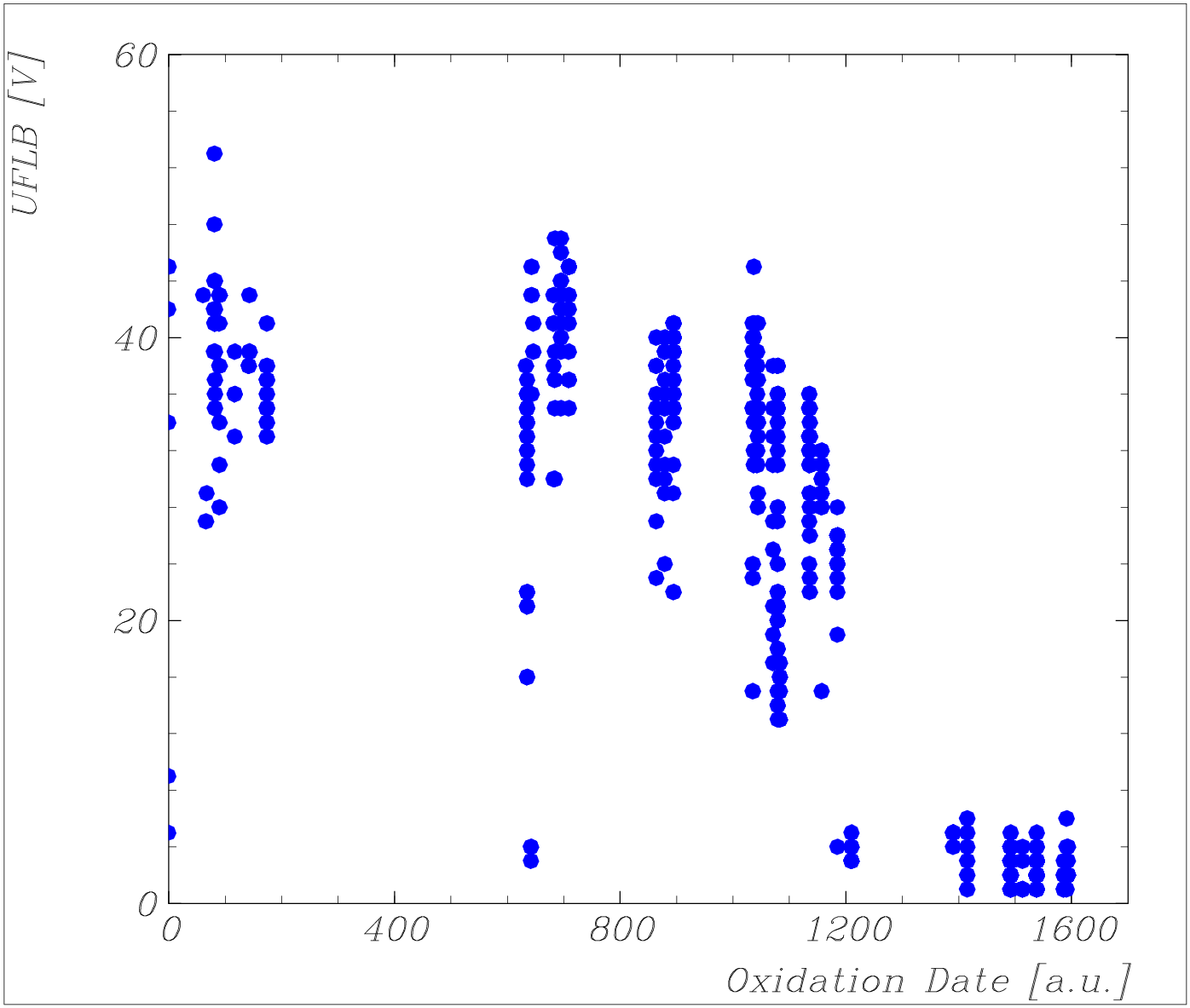,width=8cm,angle=90}
  \end{center}
  \caption[Correlation of flat band voltages with oxidation date]{%
    The correlation of the flat band voltages with the oxidation date
    is given. Newly fabricated detectors show no increase of the flat band
    voltage.}
  \label{figflbcorr}
  \begin{center}
    \leavevmode
      \epsfig{figure=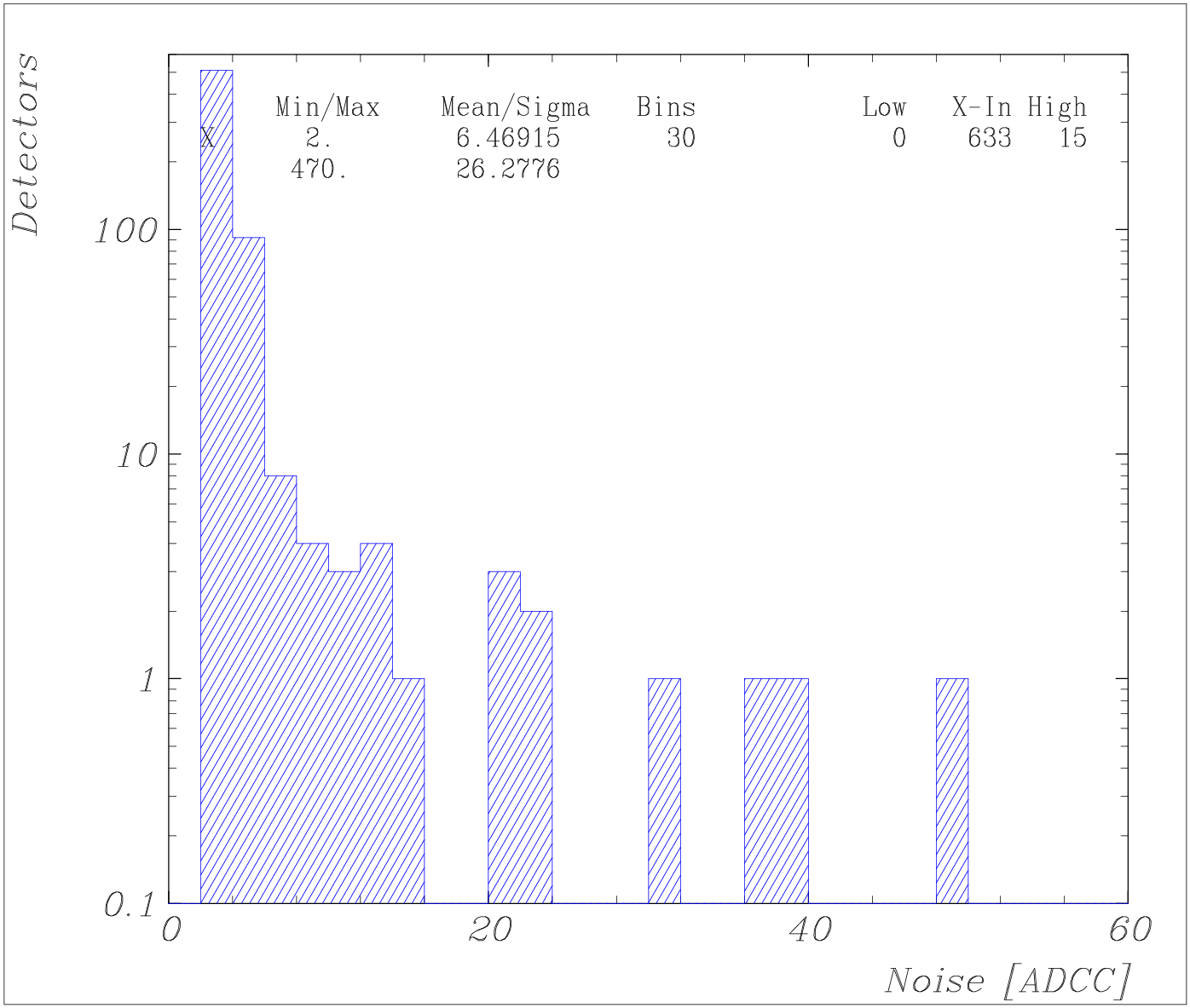,width=8cm,angle=90}
  \end{center}
  \caption[Distribution of noise of all detectors]{%
    The distribution of the noise of all detectors after the
    '94 run period is given in units of ADCC.}
  \label{fignoise}
\end{figure}

Figure \ref{figflbvo} shows the distribution of the measured flat band
voltages.  Two classes of detectors are visible. The first class has
an unchanged flat band voltage of about 2 V. The second class is
characterized by an increased flat band voltage of arround 35 V.

Figure \ref{figflbcorr} illustrates the correlation of the flat band
voltage with the oxidation date of the detector.  The correlation with
the oxidation date shown over the time period 1990-1993, during which
the detectors were produced in different oxidation batches, proves
that the newly fabricated detectors show no increase of the flat band
voltage while the older ones are characterized by a saturating value
of around 35~V.  I

\subsection{Signal Noise}

Figure \ref{fignoise} represents the noise distribution of all
detectors after the '94 run period. Most detectors do not exceed the
electronic noise of 1-2 ADC-counts (see sect.~\ref{noise}). $6\%$ of the
detectors have a noise higher than 5 ADC-counts.

\section{Discussion of Results}
\label{discu}

The investigations of the reverse current, the detector capacitance and
the signal noise of each detector used during the '94 run period leads
to the following observations~:

The reverse current is strongly increasing with time from some 100
nA's to several $\mu$A's per detector. The increase is highest for the
innermost and most backward detectors, where the electromagnetic radiation
flux impinging on the PLUG is greatest. In total $7~\%$ of the
detectors show a break-down in their I-V characteristic. These are
mainly the innermost detectors exposed to the highest radiation dose.
These detectors are no longer operational due to missing depletion of
the silicon and must therefore be replaced for the next run period.

The detector capacitance of all detectors shows an unchanged depletion
voltage, consequently radiation damage to the bulk of the silicon can
be excluded.  However the flat band voltage rises from 1-2 V as
measured in the laboratory to a saturating voltage of around 35 V
after irradiation in H1. The increase of the flat band voltage is only
observed for all detectors produced before '93, but not seen for the
latest production batches.

The noise analysis reveals that detectors, which are unstable and show
break-down, have high noise.  Hence they have a bad signal to noise
ratio and spoil the overall energy measurement of the calorimeter.
Therefore, noisy detectors have to be identified as quickly as
possible in order to exclude their signals from the readout.  In
addition detectors showing break-down affect the other detectors
included in the same bias voltage channel.  The fluctuating current of
the bad detector leads to an equivalent fluctuation of the effective
voltage for all other detectors showing up in an increased noise. To
avoid this perturbation of other detectors by damaged ones it would be
useful to have individual switches in order to be able to remove the
bias voltage for every single detector. This is foreseen for future
developments.

The main physical reason for the observed damage effect is the
generation of positively charged states in the silicon oxide
passivation layer (trapped charge) and the creation of
defects at the $SiO_{2}-Si$ interface.  Both are induced by ionizing
particles.

The interface states, which act as generation/recombination centers
for electron hole pairs (sect.~\ref{section3}), are assumed to be
responsible for the observed increase of the reverse current during
the run period. In order to prove this assumption test detectors of
much smaller size but fabricated by the same technology were
irradiated in a separate laboratory experiment with 20 keV electrons.
The reverse current was measured as function of dose in the range
between 0.5 kGy and 5 kGy. The effect is dominated by an increase of
the interface generation current with a slope of 3.6 $\mu A / cm^{2} /
kGy$. Using this result and taking into account the different field
plate areas of the test detector and the PLUG detector a current
increase of 3 $\mu A$ as observed for the '94 run period
(fig.~\ref{figDark1}) would only be consistent with an accumulated
dose in the order of 10 kGy. This large value exceeeds the results
derived from the used dosimeters by a factor of about 100. As
mentioned before (sect.  \ref{dose}), the dosimeters are only
sensitive to ionizing radiation with energies above 30 keV. Thus we
are confident that the main contribution of the radiation field
causing the observed damage effects can be attributed to low energy
photons respectively electrons \cite{mbuck95}.

Furthermore, the change of the flatband voltage induced by 20 keV
electrons was studied for devices which were kept at zero bias or at a
constant bias voltage (e.g.  80 V) during irradiation.  From
measurements as function of dose it was found that the increase of the
flat band voltage saturates at high dose. However for a bias voltage
of 80 V, which corresponds to the value typically applied to the PLUG
detectors, a saturation level of 35~V is observed at 2 kGy.  In
contrast for irradiation at zero bias saturation occurs only above 15
kGy reaching a much lower value of 10 V. This strong effect can be
explained by the different charge collection respectively trapping of
electrons and holes created in the oxide layer
\cite{pschli95},\cite{ref10}.

The result obtained for irradiation under bias are in accordance with
those observed for PLUG detectors but an estimation of the accumulated
total dose during the run period is not possible. Only a lower limit
of about 2 kGy can be stated. An open question is the different
behaviour of detectors from the latest production batches. They were
fabricated from wafers with a different oxide and show a negligible
flat band voltage shift (fig.~\ref{figflbvo}, \ref{figflbcorr}). So
far it can be stated that the radiation hardness of silicon oxides
depends strongly on the chosen technology \cite{ref11}.

An unambiguous and simple explanation for the observed break-down
effect is difficult. One possible reason is the fact that the positive
charge built up in the oxide layer especially below the field plate
region at the border of the detector causes a strong increase of the
electric field strength (fig.~\ref{detcross}). Simulations of the
field distribution have demonstrated that the field strength can
exceed values high enough for a possible avalanche break-down
\cite{ref12}, \cite{ref13}.  Inhomogeneities of the trapped oxide
charge distribution may stimulate local avalanche break-down which is
commonly called "micro plasma effect". This local effect is normally
quenched by the corresponding voltage drop at the bias resistor
resulting in large fluctuations of the detector leakage current as can
be seen in fig. \ref{figDark1}.  Such instabilities give rise to an
excessive large noise. For more detailed understanding of all damage
effects described above further studies and irradiation experiments
are in progress.

After the '94 run period $21~\%$ of the PLUG detectors were found to
have been affected by radiation damage in one way or another. As a
consequence it has been decided to reduce the number of active planes
from eight to four for the '95 run period thus providing the best
possible signal to noise ratio.  This reduction is also advantageous
because it permits better monitoring of the detector characteristics
by eliminating the need to gang two consecutive detectors (sect.
\ref{monit}).  It has also been decided for the '95 run period to
minimize the time during which the bias voltage supply is activated
thereby reducing sensitivity to radiation damage.

\section{Conclusion}

With respect to the radiation damage observed for a large amount of the
silicon detectors used in the H1-PLUG calorimeter the importance and the
difficulty for good monitoring of detector characteristics have been
pointed out. Good monitoring permits the localisation of broken-down
and noisy detectors. It is crucial for the quality of the measured
data to exclude the signals of damaged detectors from the readout.  It
would be useful to construct the bias voltage system for future
experiments in such a way, that one could switch off single damaged
detectors in a voltage channel thus avoiding a degradation in
performance of those detectors which are undamaged.

The observed radiation effects do not indicate a bulk damage of the
used silicon detectors, but a damage to the surface caused by ionizing
radiation. This result is unexpected. In particular a low radiation
hardness of the border of the used detectors forming a MOS-structure
is found, which is observed by an unexpectedly high increase of the
flat band voltage.  The possible effects of radiation damage to the
surface of silicon detectors, which will be predominantly used by the
future LHC experiments, have not been systematically investigated
especially in correlation with the applied process technology.  Given
the results presented in this paper the importance of such
investigations into surface radiation damage is stressed.

\section*{Acknowledgements}

We greatly benefited from the help of G.Lindstroem, E.Fretwurst,
H.Jarck and from the proofreading of this paper by G.Lopez. We would
like to thank H.Krueger \footnote {Student, Universit\"at Heidelberg,
  supported by the DESY Summer Student Program} for his collaboration
during the summer '94. Financial support is acknowledged from the BMFT
under contract 056HH17P.

\end{document}